\pdfoutput=1

\documentclass[11pt,twoside,a4paper,cmspaper,final,collab]{cms-tdr}
\def\svnVersion{482606P}\def\cmsCernNoTag{CERN-EP-2018-092}\def\cmsCernDate{\today}\def\cmsMessage{Published in the Journal of High Energy Physics as \href{http://dx.doi.org/10.1007/JHEP11(2018)152}{\doi{10.1007/JHEP11(2018)152.}}}

\begin{document}\cmsNoteHeader{HIG-17-007}

\hyphenation{had-ron-i-za-tion}
\hyphenation{cal-or-i-me-ter}
\hyphenation{de-vices}
\RCS$HeadURL: svn+ssh://svn.cern.ch/reps/tdr2/papers/HIG-17-007/trunk/HIG-17-007.tex $
\RCS$Id: HIG-17-007.tex 479529 2018-10-28 22:07:33Z shilpi $

\providecommand{\cmsTable}[1]{\resizebox{\textwidth}{!}{#1}}
\newcommand{\mzg}{\ensuremath{m_{\ell\ell\gamma}\xspace}}

\newlength\cmsFigWidth
\setlength\cmsFigWidth{0.4\textwidth}
\newlength\cmsTabSkip\setlength{\cmsTabSkip}{2ex}
\newlength\cmsTabSkipTwo\setlength{\cmsTabSkipTwo}{4ex}

\cmsNoteHeader{HIG-17-007}
\title{Search for the decay of a Higgs boson in the $\ell\ell\gamma$ channel in proton-proton collisions at $\sqrt{s}=13\TeV$}

\date{\today}

\abstract{
A search for a Higgs boson decaying into a pair of electrons or muons
 and a photon is described. Higgs boson decays to a $\cPZ$ boson
 and a photon
 ($\PH\to\cPZ\gamma\to\ell\ell\gamma$,$\ell=\Pe$ or $\mu$),
 or to two photons, one of which has an internal conversion
 into a muon pair
 ($\PH\to\gamma^{*}\gamma\to\mu\mu\gamma$)
 were considered. The analysis is performed using a data set recorded
 by the CMS experiment at the LHC from proton-proton collisions at a
 center-of-mass energy of 13\TeV, corresponding to an
 integrated luminosity of 35.9\fbinv. No significant
 excess above the background prediction has been found. Limits are set
 on the cross section for a standard model Higgs boson decaying to
 opposite-sign electron or muon pairs and a photon. The observed
 limits on cross section times the corresponding branching fractions
 vary between 1.4 and 4.0 (6.1 and 11.4) times the standard model cross
 section for
 $\PH\to\gamma^{*}\gamma\to\mu\mu\gamma$
 ($\PH\to\cPZ\gamma\to\ell\ell\gamma$) in the
 120--130\GeV mass range of the $\ell\ell\gamma$ system. The
 $\PH\to\gamma^*\gamma\to\mu\mu\gamma$ and
 $\PH\to\cPZ\gamma\to\ell\ell\gamma$ analyses are
 combined for $m_\PH=125\GeV$, obtaining an observed
 (expected) 95\% confidence level upper limit of 3.9\,(2.0) times the
 standard model cross section.  }

\hypersetup{%
pdfauthor={CMS Collaboration},%
pdftitle={Search for the decay of a Higgs boson in the ll gamma channel in proton-proton collisions at sqrt(s)=13 TeV},%
pdfsubject={CMS},%
pdfkeywords={CMS, Higgs boson, Z boson, photon, Dalitz}}

\maketitle

\section{Introduction}
\label{sec:intro}

Measurements of rare decays of the Higgs boson, such as
$\PH\to\gamma^*\gamma$ and $\PH\to\cPZ\gamma$, would enhance our
understanding of the standard model (SM) of particle physics, and
allow us to probe exotic couplings introduced by possible extensions
of the SM~\cite{Abba96, Chen12, Htollg-FB-Sun, Passarino}.  The decay
width can be modified by the theories involving heavy fermions,
gauge bosons or charged
scalars~\cite{Carena:2012xa, Chen:2013vi, PhysRevD.84.035027, Axion-at-LHC, Htollg-FB-Kor}.
Simple extensions of the SM
like two Higgs doublet models, or the minimal supersymmetric
standard model also exhibit similar
features~\cite{Chiang:2012qz}. Certain coefficients of the
dimension-6 extension of the standard model effective field theory
 can be constrained by measuring the $\PH\to\cPZ\gamma$
branching ratio precisely~\cite{Dawson:2018pyl}.
As an example, a model~\cite{Chiang:2012qz} which includes a hypercharge zero triplet extension,
shows a modification in $\mathcal{B}(\PH\to\cPZ\gamma)$, with respect to the SM value,
of about 10\% for an additional scalar field with mass between 0 and 400\GeV.

\begin{figure*}[b]
  \centering {\includegraphics[width=0.23\textwidth]{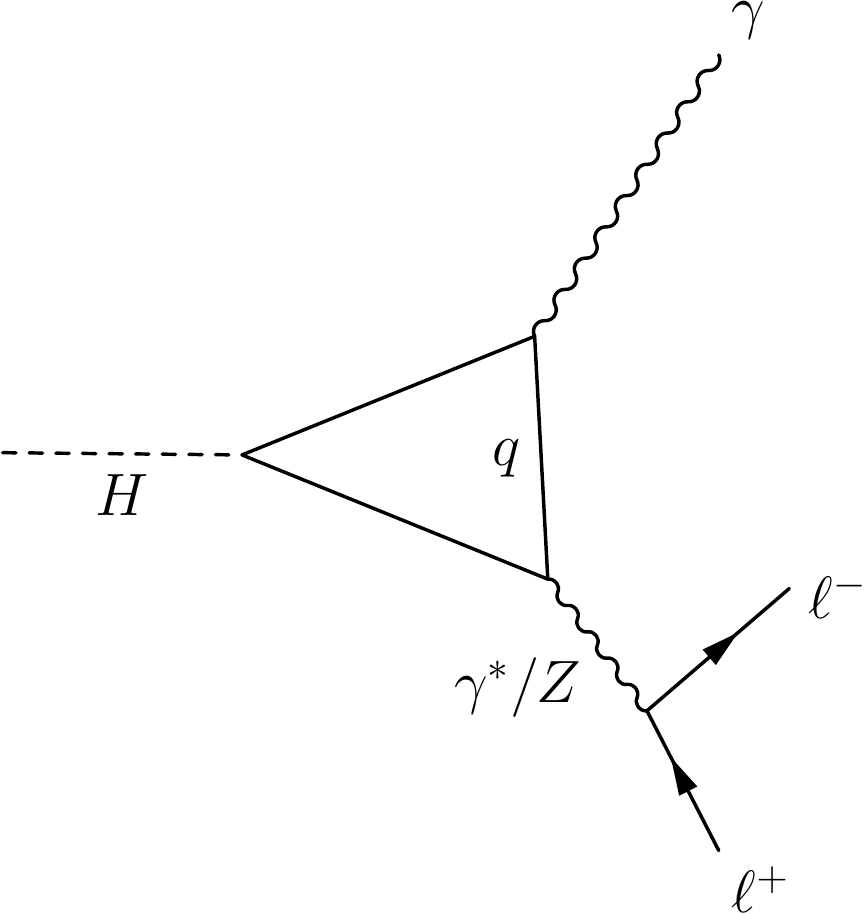}}\hfil
    {\includegraphics[width=0.23\textwidth]{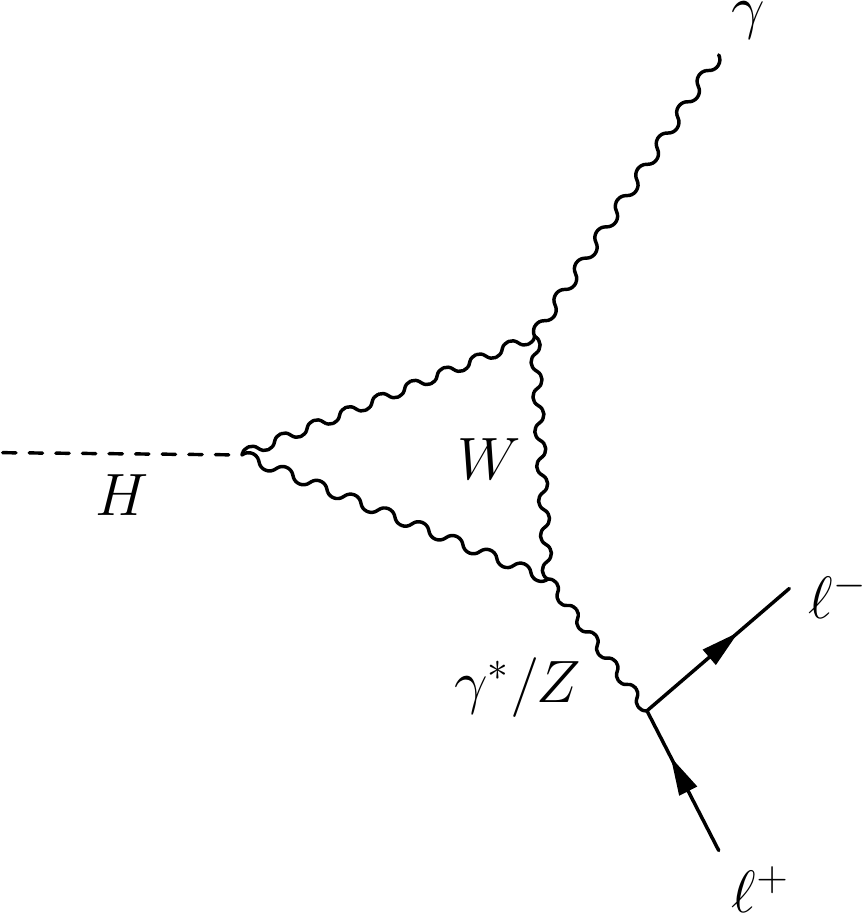}}\hfil
    {\includegraphics[width=0.23\textwidth]{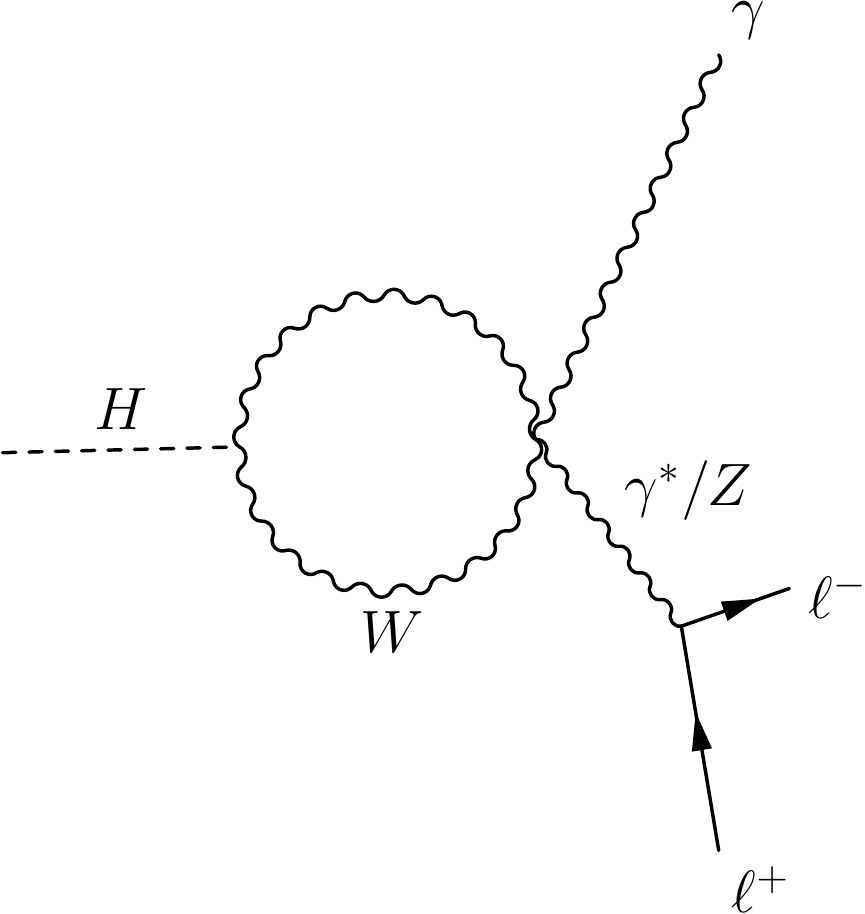}}
    {\includegraphics[width=0.23\textwidth]{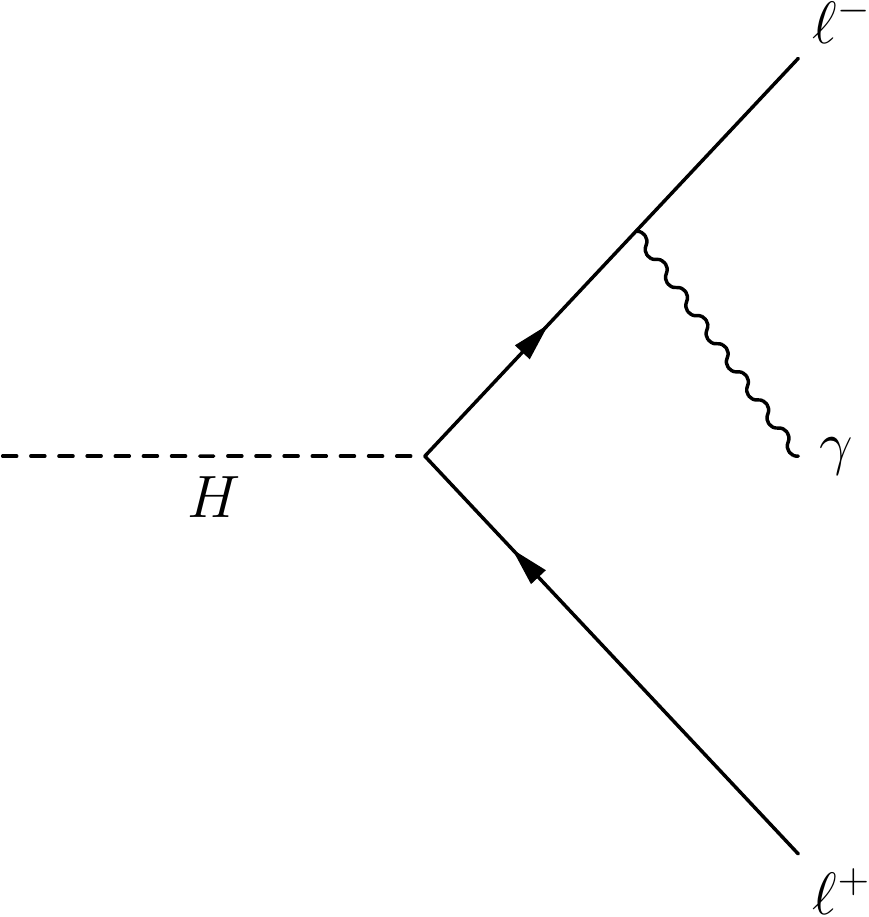}} \hfil
    {\includegraphics[width=0.23\textwidth]{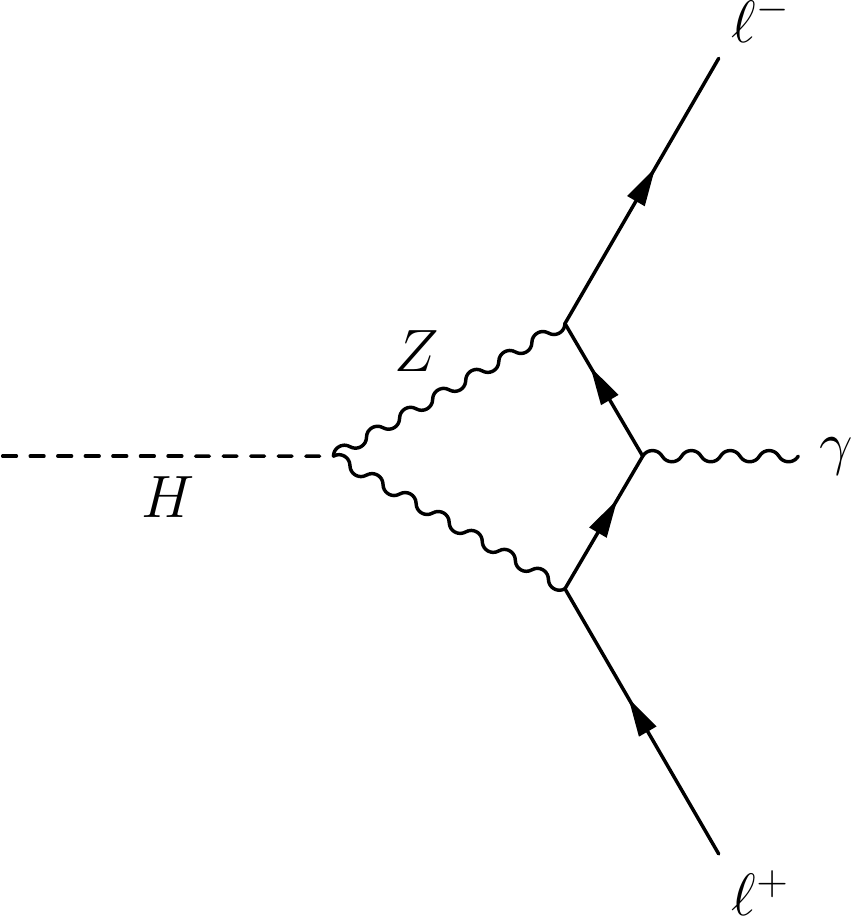}} \hfil
    {\includegraphics[width=0.23\textwidth]{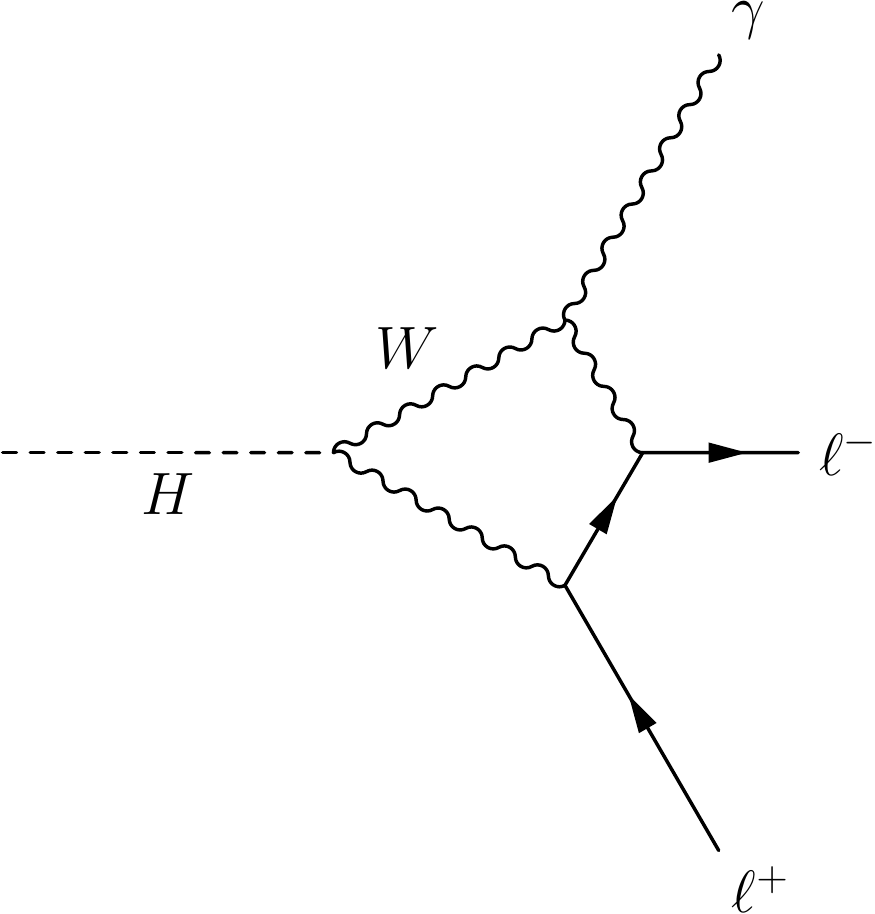}} \hfil
    {\includegraphics[width=0.23\textwidth]{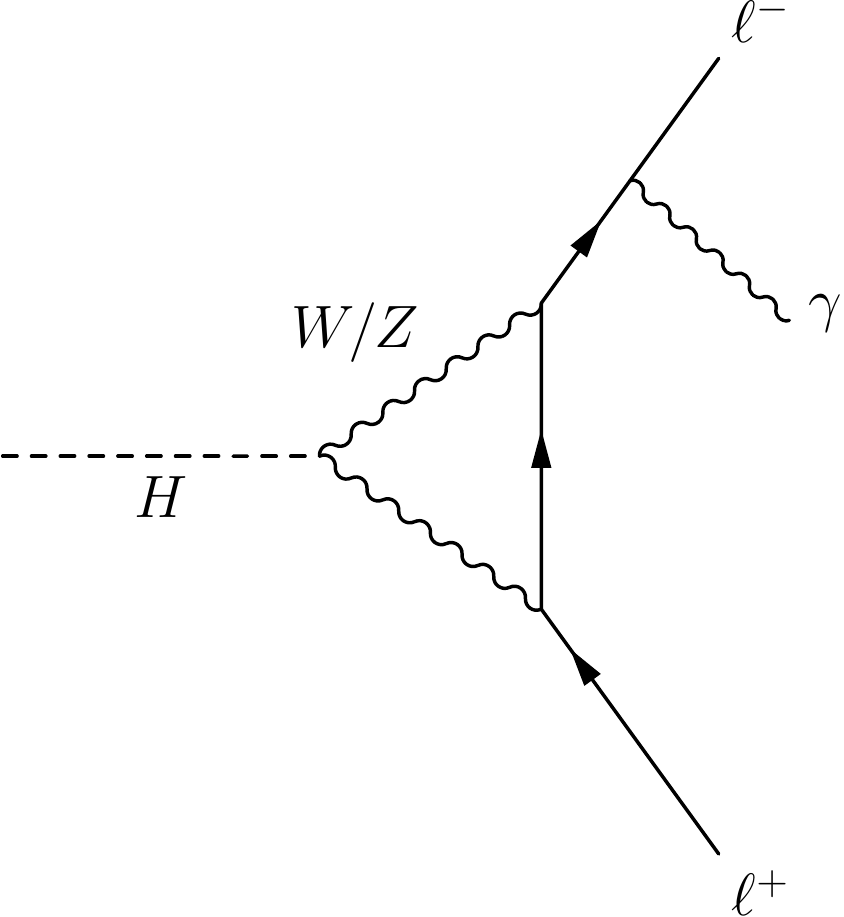}} \caption{Dominant
    Feynman diagrams contributing to the $\PH\to\ell\ell\gamma$
    process.  } \label{fig:fey}
\end{figure*}

In the search for $\PH\to\gamma^*\gamma\to\ell\ell\gamma$, the leptonic channel,
$\gamma^*/\cPZ \to \ell\ell$ ($\ell=\Pe$ or $\mu$) is most promising as it
has relatively low background. The diagrams in Fig.~\ref{fig:fey}
illustrate the dominant Higgs boson decay channels contributing to
these final states.  The $\PH\to\gamma^*\gamma\to\ell\ell\gamma$
and $\PH\to\cPZ\gamma\to\ell\ell\gamma$ diagrams correspond to the same
initial and final state and interfere with each other.
Experimentally one can separate the  off- and on-shell
contributions, and define the respective signal regions, using a selection
based on the invariant mass of the dilepton
system, $m_{\ell\ell} = m_{\gamma^*/\cPZ}$. For the measurements presented in this paper
a threshold of   $m_{\ell\ell} = 50\GeV$ is used to
separate the two processes.

It is informative to express the branching fractions for these decays
relative to the $\PH\to\gamma\gamma$ process. In the SM, for a Higgs boson
with mass $m_\PH = 125\GeV$~\cite{Aad:2015zhl, bib:htozz2016}, these ratios are:
\begin{equation}
  \frac{\mathcal{B}(\PH\to\gamma^*\gamma \to \mu\mu\gamma)}{\mathcal{B}(\PH\to\gamma\gamma)} = (1.69\pm0.10)\%, \quad
  \frac{\mathcal{B}(\PH\to\cPZ\gamma\to \Pep\Pem\gamma/\mu\mu\gamma)}{\mathcal{B}(\PH\to\gamma\gamma)}  = (2.27\pm0.14)\%,
  \label{eq:fraction}
\end{equation}
where $\mathcal{B}(\PH\to\cPZ\gamma\to\Pep\Pem\gamma/\mu\mu\gamma) = 0.051\times10^{-3}$ and
$\mathcal{B}(\PH\to\gamma\gamma) = 2.27\times10^{-3}$ are taken from
Ref.~\cite{LHC-YR4}, and $\mathcal{B}(\PH\to\gamma^*\gamma \to
\mu\mu\gamma) = 3.83\times10^{-5}$ is obtained with the \MCFM 7.0.1 program~\cite{MCFM7},
which is in agreement with calculations in
Refs.~\cite{Dicus13,Dicus14,Firan07}.

The ATLAS and CMS Collaborations at the CERN LHC have both performed
searches for the decay $\PH\to\cPZ\gamma\to\ell\ell\gamma$~\cite{atl-HZG,cms-HZG} at $\sqrt{s}=7$ and
$8\TeV$. The ATLAS Collaboration set an upper limit on
$\sigma/\sigma_\mathrm{SM}$ of 11 (where $\sigma_\mathrm{SM}$ is the
expected cross section of the SM signal process) at 95\% confidence
level (\CL) for an SM Higgs boson with $m_\PH=125.5\GeV$, and the CMS
Collaboration set an upper limit of 9.5 at 95\% \CL for
$m_\PH=125\GeV$.  The CMS Collaboration has also searched for the
$\PH\to\gamma^*\gamma\to\ell\ell\gamma$ process with $m_{\ell\ell} <
20$ (1.5)\GeV in the dimuon (dielectron) channel at
$8\TeV$~\cite{cms-Dalitz}. The two channels were combined to set an upper limit
of 6.7 at 95\% \CL on $\sigma/\sigma_\mathrm{SM}$ for $m_\PH=125\GeV$.
The ATLAS Collaboration has also performed a search for
$\PH\to\cPZ\gamma\to\ell\ell\gamma$ at $\sqrt{s}=13\TeV$
 using 36.1\fbinv of data collected in 2016. This search set an
upper limit on $\sigma/\sigma_\mathrm{SM}$ of 6.6 at 95\% \CL for an SM
Higgs boson with $m_\PH=125.09\GeV$~\cite{Aaboud:2017uhw}.

This paper describes a search for Higgs bosons decaying to
 $\PH\to\gamma^*\gamma\to\mu\mu\gamma$ and
 $\PH\to\cPZ\gamma\to\ell\ell\gamma$ at 13\TeV. The study
of the $\PH\to\gamma^*\gamma\to\Pe\Pe\gamma$ decay is challenging~\cite{cms-Dalitz},
because if $m_{\ell\ell}$ is low, the pair of electron showers merge in the
electromagnetic calorimeter (ECAL). This merging makes it difficult to
trigger on such events and also to reconstruct them offline.
Therefore, this channel is not included in the present analysis.

The analysis uses a data sample of proton-proton ($\Pp\Pp$) collisions
at a center-of-mass energy of 13\TeV recorded by the CMS experiment
during 2016, corresponding to an integrated luminosity of 35.9\fbinv.
The sensitivity of the search is enhanced by dividing the selected
events into mutually exclusive classes, according to the expected mass
resolution and the signal-to-background ratio, and then combining the
results from each class.  This paper is structured as follows. In
Section~\ref{sec:cms}, the CMS detector is described. The event
selection used in the analysis is outlined in
Section~\ref{sec:preselection}. Section~\ref{sec:modeling} discusses about
 signal and background modeling.
Systematic uncertainties and the results of this study are
presented in Section~\ref{sec:results}, followed by the summary in
Section~\ref{sec:summary}.

\section{The CMS detector and trigger}
\label{sec:cms}
 A detailed description of the CMS detector can be found in
Ref.~\cite{CMS-Jinst}. The central feature of the CMS apparatus is a
superconducting solenoid, 13\unit{m} in length and 6\unit{m} in
diameter, which provides an axial magnetic field of 3.8\unit{T}.
Within the field volume there are several particle detection
systems. Charged-particle trajectories are measured by silicon pixel
and silicon strip trackers, covering $0\le \phi \le 2\pi$ in azimuth
and $\abs{\eta} < 2.5$ in pseudorapidity.  A lead-tungstate
crystal ECAL and a brass and
scintillator hadron calorimeter (HCAL) surround the tracking volume
and cover the region $\abs{\eta} < 3$. They provide energy
measurements of photons, electrons and hadronic jets.  The ECAL is
partitioned into a barrel region with $\abs{\eta} < 1.48$ and two endcaps
that extend up to $\abs{\eta} = 3$.  A lead and silicon-strip preshower
detector is located in front of the endcap of the ECAL.
Muons are identified and measured in gas-ionization
detectors embedded in the steel return yoke outside the solenoid. The
detector is nearly hermetic, allowing energy balance measurements in
the plane transverse to the beam direction.

A two-level trigger system selects collision events of interest for
physics analysis ~\cite{Khachatryan:2016bia}.  The trigger used in the
$\PH\to\gamma^*\gamma\to\mu\mu\gamma$ channel requires a muon and a
photon with transverse momenta, \pt, greater than 17 and 30\GeV,
respectively. The trigger efficiency is determined using signal events
in simulation and $\mu\mu\gamma$ events in data using an orthogonal
data set selected with a single muon trigger.  For events satisfying
the selection criteria described in Section~\ref{sec:preselection} the
trigger efficiency is 83\% in both cases. The
$\PH\to\cPZ\gamma\to\ell\ell\gamma$ events are required to pass at
least one of the dielectron or dimuon triggers. The dielectron trigger
requires a leading (subleading) electron with \pt greater than
23\,(12)\GeV. The dimuon trigger requires a leading (subleading) muon
with \pt greater than 17\,(8)\GeV. The efficiencies of these dilepton
triggers as measured in data, for events satisfying the selection
criteria, are dependent on the $\pt$ and $\eta$ of the leptons and are
measured to be 90--98\% and 93--95\% for the $\Pe\Pe\gamma$ and
$\mu\mu\gamma$ channels, respectively.

\section{Event selection}
\label{sec:preselection}
Selected events are required to have at least one good primary vertex,
with reconstructed longitudinal position within 24\unit{cm} of the
geometric center of the detector and transverse position within
2\unit{cm} of the beam interaction region.  Due to the high
instantaneous luminosity of the LHC, there are multiple $\Pp\Pp$
interactions per bunch crossing (pileup). In the case of multiple
vertices, the vertex with the largest value of summed physics-object
$\pt^2$ is taken to be the primary $\Pp\Pp$ interaction vertex.  The
physics objects chosen are those that have been defined using
information from the tracking detector, including jets, the associated
missing transverse momentum, which is defined as the negative vector
sum of the \pt of those jets, and charged leptons.  All leptons, which
are used to select events, are required to have transverse and
longitudinal impact parameters with respect to the primary vertex
smaller than 5 and 10\,mm, respectively.

The particle-flow (PF) event reconstruction algorithm~\cite{PF2017} is used to reconstruct and identify each individual particle using an optimized combination of information from the various elements of the CMS detector.

Photon candidates are reconstructed from clusters of crystals in the
ECAL with significant energy deposits \cite{CMS-EGM-14-001}.  Clusters
are grouped into superclusters to recover the energy from electron
bremsstrahlung and photons converting in the tracker.  In the endcaps,
the preshower detector energy is also included for the region covered
by the preshower detector ($1.65 < \abs{\eta} <2.6$).  The
clustering algorithms result in almost complete recovery of the energy
of photons. Photon candidates are selected with a multivariate
discriminant that uses, as inputs, isolation variables, the ratio of
the energy in the HCAL behind an electromagnetic
supercluster to the supercluster energy, and the transverse width of
the electromagnetic shower. Isolation variables are based on particle
candidates from the PF algorithm.  A conversion-safe electron
veto \cite{CMS-EGM-14-001} is applied to avoid misidentifying an
electron as a photon. This vetoes events that have a charged particle
track with a hit in the inner layer of the pixel detector that points
to the photon cluster in the ECAL, unless that track is matched to a
conversion vertex.  Photons are required to lie in the geometrical
region $\abs{\eta}<2.5$ and have $\pt>15\GeV$. The efficiency of the
photon identification is measured from $\cPZ\to\Pe\Pe$ events using
tag-and-probe techniques~\cite{cite:tagandprobe}. It is found to be
between 84 and 91 (77 and 94)\% in the barrel (endcaps) depending on
the $\pt$ of the photon, after including the electron veto
inefficiencies measured with $\cPZ\to\mu\mu\gamma$ events, where the
photon is produced by final-state radiation.

Electron reconstruction starts from superclusters in the ECAL, which
are matched to hits in the silicon strip and the pixel detectors.  The
energy of electrons is determined from a combination of the electron
track momentum at the main interaction vertex and the energy of the
corresponding ECAL cluster.  Electrons are selected using a
multivariate discriminant that includes observables sensitive to the
presence of bremsstrahlung along the electron trajectory, the
geometrical and momentum-energy matching between the electron
trajectory and the energy of the associated cluster in the ECAL, the
shape of the electromagnetic shower in the ECAL, and the variables
that discriminate against electrons originating from photon
conversions \cite{bib:htozz2016}. In this analysis, we accept
electrons with $\pt>7\GeV$ and $\abs{\eta}<2.5$.

Muon candidates are reconstructed in the tracker and identified by the
PF algorithm using hits in the tracker and the muon systems. The
matching between the inner and outer tracks proceeds either
outside-in, starting from a track in the muon system, or inside-out,
starting from a track in the silicon tracker. In the latter case,
tracks that match track segments in only one or two planes of the muon
system are also considered in the analysis in order to collect very
low-$\pt$ muons that may not have sufficient energy to penetrate the
entire muon system. The muons are selected from the reconstructed muon
track candidates by applying minimal requirements on the track in both
the muon system and inner tracker system, and taking into account
compatibility with small energy deposits in the calorimeters. We
accept muons with $\pt>4$\GeV and $\abs{\eta}<2.4$ \cite{bib:htozz2016}.

The relative isolation variable, used to select prompt leptons,
is defined as:
\begin{equation}
\label{eqn:pfiso}
{\cal I}^{\ell} \equiv \Big( \sum \PT^\text{charged} +
                                 \max\big[ 0, \sum \PT^\text{neutral}
                                 +
                                  \sum \PT^{\Pgg}
                                 - \PT^\mathrm{PU}(\ell) \big] \Big)
                                 / \PT^{\ell},
\end{equation}
 and is required to be less than 0.35, where
$\sum \PT^\text{charged}$ is the scalar sum of the transverse momenta
of charged hadrons originating from the primary vertex,
$\sum \PT^\text{neutral}$ and $\sum \PT^{\Pgg}$ are the scalar sums of
the transverse momenta for neutral hadrons and photons, respectively,
and $\sum \PT^\mathrm{PU}(\ell)$ accounts for the contribution of
neutral pileup particles.  The isolation sums are performed over a
cone of angular radius $\DR = \sqrt{\smash[b]{(\Delta\phi)^2 +
(\Delta\eta)^2}} = 0.3$ around the lepton direction at the primary
vertex. For muons, $\PT^\mathrm{PU}(\Pgm) \equiv
0.5 \, \sum_i \PT^{\mathrm{PU}, i}$, where $i$ runs over the momenta
of the charged hadron PF candidates not originating from the primary
vertex. For electrons, $\PT^\mathrm{PU}(\Pe) \equiv \rho \,
A_\text{eff}$, where the effective area $ A_\text{eff}$ is a
coefficient that is dependent on electron $\eta$ and is chosen in such
a way that the isolation efficiency is independent of pileup (PU), and
$\rho$ is the median of the $\PT$ density distribution for neutral
particles~\cite{fastjet,Cacciari:2007fd,Cacciari:2008gn}.  Finally,
$\PT^{\ell}$ is the transverse momentum of the selected lepton.  To
suppress muons originating from non-prompt decays of hadrons and
electrons from photon conversions, we require each lepton track to
have a 3D impact parameter with respect to the primary vertex that is
less than four times its uncertainty.

The optimized electron selection criteria, including the isolation
requirement, give an efficiency of approximately 85--93 (81--92)\%
in the barrel (endcaps) for electrons from $\PW$ or $\cPZ$ bosons.
For muons, the identification is tuned to maintain efficiency at low
$\DR$ where the two muons are close to each other. The
identification and isolation efficiency for single muons from
$\cPZ\to\mu\mu$ or $\cPJgy$ meson decays is 85--97 (88--96)\%
in the barrel (endcaps). In the case of the
$\PH\to\gamma^*\gamma\to\mu\mu\gamma$, the $\DR(\mu\mu)$ between the
two muons is small due to their low invariant mass and the high $\PT$
of the $\gamma^*$. Hence, no isolation requirement is applied to the
subleading muons as they are within the isolation cone of the leading
muons in most events. Also, if the subleading muon falls within the isolation cone of 
the leading muon, it is not included in the calculation of the isolation variable. 
The identification efficiency of muons from
$\gamma^*$ is approximately 94--98 (92--97)\% in the barrel
(endcaps).

Selected events are classified as described in detail below.  The
dijet-tagged (explained in Section~\ref{sec:dalitz}) event class uses
jets that are built by clustering the PF candidates using the
anti-\kt clustering algorithm with a distance parameter of 0.4
using the \FASTJET software package~\cite{fastjet}.  Charged
PF candidates from pileup vertices are discarded to reduce the
contribution to the jet energies from pileup interactions. An offset
correction is applied to account for the remaining contributions.  In
situ measurements of the momentum balance in dijet, photon+jet, $\cPZ$+jet,
and multijet events are used to account for any residual differences
in jet energy scale in data and simulation.  Additional selection
criteria are applied to each event to remove spurious jet-like
features originating from isolated noise patterns in certain HCAL
regions. Calibrated and corrected jets are required to have $\pt$
greater than 30\GeV and $\abs{\eta}<4.7$, and to be separated by
at least 0.4 in $\DR$ from leptons and photons passing the selection
requirements described above.

\subsection{\texorpdfstring{$\PH\to\gamma^*\gamma\to\mu\mu\gamma$}{Higgs to gamma* gamma to mu mu gamma} selection}
\label{sec:dalitz}

In the $\PH\to\gamma^*\gamma\to\mu\mu\gamma$ search we select events
with two muons and a photon, where the muons must have opposite
charges and $\PT > 20\,(4)\GeV$ for the leading (subleading) muon.
The \PT requirement on the leading muon is driven by the trigger
threshold, and that on the subleading muon by the minimum energy
needed to reach the muon system.  The photon and dimuon transverse
momenta both must satisfy $\pt > 0.30m_{\mu\mu\gamma}$, where
$m_{\mu\mu\gamma}$ is the invariant mass of the $\mu\mu\gamma$ system.
 This requirement rejects the $\gamma^*+$jet and $\gamma+$jet backgrounds without
any loss in the signal sensitivity and without introducing a bias in the
$m_{\mu\mu\gamma}$ spectrum.
The separation between each muon and the photon is required to satisfy
$\DR>1$ in order to suppress Drell--Yan background events with
final-state radiation.

The dimuon invariant mass is required to be less than 50\GeV to make
this selection and the $\cPZ\gamma$ selection described in
Section~\ref{sec:HtoZGsel} mutually exclusive. Events with a dimuon
mass in the ranges $2.9< m_{\mu\mu}< 3.3\GeV$ and $9.3 < m_{\mu\mu} <
9.7\GeV$ are rejected to avoid $\cPJgy\to\mu\mu$ and
$\Upsilon(nS)\to\mu\mu$ contamination, respectively. The invariant
mass $m_{\mu\mu\gamma}$ is required to satisfy $110 < m_{\mu\mu\gamma}
<170\GeV$.  In the cases where there are multiple dilepton pairs in
the event, the one with the smallest dimuon invariant mass is chosen.

A variable $\RNINE$ is defined as the energy sum of the 3$\times$3
ECAL crystals centered on the most energetic crystal in the
supercluster divided by the energy of the supercluster. The selected
events are separated into four mutually exclusive event classes based
on the $\RNINE$ and $\eta$ of the photon and the presence of jets. An
$\RNINE$ value of 0.94 is used to separate the reconstructed photons
into two regions. The region containing unconverted photons, with
larger values of $\RNINE$ and better energy resolution, has a smaller
background.  By separating events into two regions of low/high
$\RNINE$ value, the sensitivity of the analysis is increased.  We
therefore have the following four categories: events that require the
presence of at least two jets passing the selection criteria as
described below; photon in the ECAL barrel (EB) region with a high
$\RNINE$ value; photon in the barrel with low $\RNINE$ value; and
photon in the ECAL endcap (EE) regions. Only events that do not pass
the dijet tag are included in the EB or EE classes. By using this
event classification scheme, as opposed to combining all events into
one class, the sensitivity of this analysis is increased by 11\%.

For the dijet tag event class the two highest transverse energy jets
are used and the requirements are: (i) the difference in
pseudorapidity between the two jets is greater than 3.5; (ii) the
Zeppenfeld variable~\cite{Rainwater:1996ud} ($\eta_{\ell\ell\gamma} -
(\eta_{\mathrm{j1}}+\eta_{\mathrm{j2}})/2$) is less than 2.5, where
$\eta_{\ell\ell\gamma}$ is the $\eta$ of the $\ell\ell\gamma$ system
and $\eta_{\mathrm{j1}}$ and $\eta_\mathrm{{j2}}$ are the pseudorapidities of the
leading and subleading jets, respectively; (iii) the dijet mass is
greater than 500\GeV; and (iv) the difference in azimuthal angles
between the dijet system and the $\ell\ell\gamma$ system is greater
than 2.4.  These requirements mainly target the vector boson fusion
(VBF) production mechanism of the Higgs boson.

The resulting acceptance times efficiency for
$\Pp\Pp\to\PH\to \gamma^{*}\gamma\to\mu\mu\gamma$ is 26--27\% for
$m_{\PH}$ between 120 and 130\GeV.

\subsection{\texorpdfstring{$\PH\to\cPZ\gamma\to\ell\ell\gamma$}{Higgs to Z gamma to ell ell gamma}  selection}

\label{sec:HtoZGsel}
In the $\PH\to\cPZ\gamma\to\ell\ell\gamma$ search, events with a
photon and with at least two same-flavor leptons ($\Pe$ or $\mu$)
consistent with a $\cPZ$ boson decay are selected. All particles must
be isolated, and have {$\pt$} greater than 25\,(15)\GeV for the
leading\,(subleading) electron, 20\,(10)\GeV for the
leading\,(subleading) muon, and 15\GeV for the photon. In the cases
where there are multiple dilepton pairs in the event, the one with the
mass closest to the $\cPZ$ boson nominal mass ~\cite{Patrignani:2016xqp} is
selected. The invariant mass of the selected pair is required to be
larger than 50\GeV. This ensures that the
$\PH\to\cPZ\gamma\to\ell\ell\gamma$ event selection is orthogonal to
that for $\PH\to\gamma^*\gamma\to\mu\mu\gamma$.

The events are required to have a photon with $\ET > 0.14m_{\ell\ell\gamma}$,
which rejects the $\cPZ+$jets background without significant
loss in signal sensitivity and without introducing a bias in the
$m_{\ell\ell\gamma}$ spectrum. Leptons are required to have $\DR>0.4$
with respect to the photon in order to reject events with final-state
radiation.  In addition, we require $m_{\ell\ell\gamma} + m_{\ell\ell}
> 185\GeV$ to reject events with final-state radiation from Drell--Yan
processes. Finally, the invariant mass of the $\ell\ell\gamma$ system
is required to be $115<m_{\ell\ell\gamma}<170\GeV$.

The selected events are classified into mutually exclusive categories.
A lepton-tag class contains events with an additional electron (or
muon) with $\pt>7$\,(5)\GeV, to target Higgs boson production in
association with either a $\cPZ$ or $\PW$ boson.  Events not included in the
lepton class are considered for the dijet class. In this case the
criteria described in Section~\ref{sec:dalitz} are used to select
events containing a dijet, targeting Higgs boson production in a VBF
process.  The next class considered is the boosted class, which
requires that the $\pt$ of the $\ell\ell\gamma$ system is greater than
60\GeV in order to enhance the fraction of events that contain a
Lorentz-boosted Higgs boson recoiling against a jet. Events that do
not fall into these three classes are placed in the untagged
categories.  A significant fraction of the signal events are expected
to have the photon and both leptons in the barrel, while only a sixth
of the signal events have the photon in the endcap. This is in
contrast to the background, where about one third of the events are
expected to have a photon in the endcap. Furthermore, events where the
photon does not convert to $\Pe^+\Pe^-$ have a smaller fraction of
background events and better energy resolution. For these reasons, the
untagged events are classified into four categories according to the
pseudorapidity of the leptons and photon, and the $\RNINE$ value of
the photon.  These categories are indicated as untagged 1, untagged 2,
untagged 3 and untagged 4 as shown in Table~\ref{tab:untag}.

\begin{table*}[htb!]
\centering
\topcaption{Categories in $\PH\to\cPZ\gamma\to\ell\ell\gamma$ search. The electron and muon channels are considered separately in all classes except for the lepton-tag class.}
\label{tab:untag}
\cmsTable{
\begin{tabular}{ccc}
 Category  & $\Pep\Pem\gamma$ & $\mu^+\mu^-\gamma$ \\ \hline
 Lepton tag
           & \multicolumn{2}{c}{Additional electron ($\pt>7\GeV$) or muon ($\pt>5\GeV$)} \\[\cmsTabSkip]
 \multirow{1}{*}{Dijet tag}
           & At least 2 jets required   &  At least 2 jets required    \\
           & dijet selection (Section~\ref{sec:dalitz})             & dijet selection (Section~\ref{sec:dalitz}) \\[\cmsTabSkip]
  \multirow{1}{*}{Boosted}
            & $\pt(\Pe\Pe\gamma)>60\GeV$ & $\pt(\mu\mu\gamma)>60\GeV$ \\[\cmsTabSkip]
  \multirow{4}{*}{Untagged 1}
  & Photon $ 0<\abs{\eta}<1.4442$ & Photon $0<\abs{\eta}<1.4442$\\
                      & Both leptons $0<\abs{\eta}<1.4442$ & Both leptons $0<\abs{\eta}<2.1$\\
                      & $\RNINE>0.94$ & and one lepton $0<\abs{\eta}<0.9$\\
                      &  & $\RNINE>0.94$ \\[\cmsTabSkip]
  \multirow{4}{*}{Untagged 2}
  & Photon $ 0<\abs{\eta}<1.4442$ & Photon $0<\abs{\eta}<1.4442$\\
                      & Both leptons $0<\abs{\eta}<1.4442$ & Both leptons $0<\abs{\eta}<2.1$\\
                      & $\RNINE<0.94$ & and one lepton $0<\abs{\eta}<0.9$\\
                      &  & $\RNINE<0.94$ \\[\cmsTabSkip]
  \multirow{4}{*}{Untagged 3}
  & Photon $0<\abs{\eta}<1.4442$ & Photon $0<\abs{\eta}<1.4442$\\
                      & At least one lepton $1.4442 <\abs{\eta}<2.5$ & Both leptons in $\abs{\eta}>0.9$ \\
                      & No requirement on $\RNINE$ & or one lepton in $2.1<\abs{\eta}<2.4$\\
                      &  & No requirement on $\RNINE$ \\[\cmsTabSkip]
  \multirow{3}{*}{Untagged 4}
  & Photon $ 1.566<\abs{\eta}<2.5$  & Photon $1.566<\abs{\eta}<2.5$ \\
                      & Both leptons $0<\abs{\eta}<2.5$ & Both leptons $0<\abs{\eta}<2.4$\\
                      & No requirement on $\RNINE$  & No requirement on $\RNINE$ \\
 \end{tabular}
}
\end{table*}
It should be noted that the electron and muon channels are considered
separately in all classes except for the lepton-tag class where the
number of events is small.  This event classification scheme increases
the sensitivity of the analysis by 18\%.  The resulting acceptance
times efficiency for $\Pp\Pp\to\PH\to \cPZ\gamma\to\ell\ell\gamma$ in
the electron (muon) channel is between 18 and 24 (25 and 31)\% for
$m_{\PH}$ between 120 and 130\GeV.

A complete list of all the categories considered in the analysis
($\Pp\Pp\to\PH\to\gamma^*\gamma\to\mu\mu\gamma$ and
$\Pp\Pp\to\PH\to \cPZ\gamma\to\ell\ell\gamma$), together with the
expected yields for a 125\GeV SM Higgs boson signal processes, is
shown in Table~\ref{tab:yield}. This table also reports yields from
signal processes: gluon-gluon fusion ($\Pg\Pg\PH$), vector boson fusion (VBF),
associated $\mathrm{V}\PH$ production ($\mathrm{V}\PH$) and Higgs boson production in
association with top quarks ($\ttbar\PH$).

\begin{table*}[bth]
\centering
\topcaption{
Expected signal yields for a 125\GeV SM Higgs boson,
corresponding to an integrated luminosity of 35.9\fbinv,
for all categories in the
$\PH\to\gamma^*\gamma\to\mu\mu\gamma$ and
$\PH\to\cPZ\gamma\to\ell\ell\gamma$ processes in the narrowest $\ell\ell\gamma$
invariant mass window around 125\GeV containing 68.3\% of the
expected signal distribution.
  \label{tab:yield}}
\begin{tabular}{cccccc}

  &  &  & \multicolumn{3}{c}{Number of signal events} \\
Analysis  &    Channel     &     Category        & \multicolumn{3}{c}{for $m_\PH=125\GeV$} \\
  &             &  &       $\Pg\Pg\PH$    &  VBF & $\mathrm{V}\PH+\ttbar\PH$ \\
\hline
\multirow{4}{*}{$\PH\to\gamma^*\gamma\to\mu\mu\gamma$}
& $\mu\mu$ & EB, high $\RNINE$ & 9.18  & 0.47  &  0.33 \\
& $\mu\mu$ & EB, low $\RNINE$ & 5.17  & 0.27  &   0.18 \\
& $\mu\mu$ & EE               & 3.80  & 0.20  &   0.25 \\
& $\mu\mu$ & Dijet tag             & 0.45  & 0.39  &  0.01 \\\\[\cmsTabSkipTwo]
\multirow{13}{*}{$\PH\to\cPZ\gamma\to\ell\ell\gamma$}
& $\Pe\Pe+\mu\mu$& Lepton tag &  0.08& 0.014& 0.33\\
& $\Pe\Pe$& Dijet tag & 0.34 & 0.47 & 0.02 \\
& $\Pe\Pe$& Boosted  &  3.38 & 0.56 & 0.33\\
& $\Pe\Pe$& Untagged 1& 5.2 & 0.15 &  0.06\\
& $\Pe\Pe$& Untagged 2& 3.2 & 0.09 &  0.04\\
& $\Pe\Pe$& Untagged 3&  3.9 &  0.12 & 0.06\\
& $\Pe\Pe$& Untagged 4& 2.8 & 0.08 &  0.04\\
& $\mu\mu$& Dijet tag &  0.44 & 0.62 & 0.02\\
& $\mu\mu$& Boosted  &  4.51 & 0.74 & 0.44\\
 & $\mu\mu$&Untagged 1&  7.6 & 0.22  & 0.097\\
 & $\mu\mu$&Untagged 2&  4.8 & 0.14  & 0.06\\
 & $\mu\mu$&Untagged 3&  4.1 & 0.12  & 0.06\\
 & $\mu\mu$&Untagged 4&  3.5 & 0.11  & 0.06\\
\end{tabular}
\end{table*}

\section{Signal and background modeling}
\label{sec:modeling}
The search for signal events is performed using a shape-based analysis
of $\ell\ell\gamma$ invariant mass distributions. The background is
estimated from data and the signal is estimated using the simulation.
Even though the background is estimated from data, simulated samples
are used in the $\PH\to\cPZ\gamma\to\ell\ell\gamma$ search to optimize
the event classes.  The main background, $\Pp\Pp\to Z\gamma$,
is generated at next-to-leading order (NLO) using the \MGvATNLO
generator~\cite{Alwall:2014hca}. The $\PZ(\ell\ell)$+jets events with
a jet misidentified as a photon are another important source of
background and are generated at NLO using \MGvATNLO.  The NLO parton
distribution function\,(PDF) set, NNPDF3.0~\cite{nnpdf30}, and the
CUETP8M1~\cite{Khachatryan:2015pea} underlying event tune are used to
generate these samples.  All background events are interfaced
with \PYTHIA 8.205~\cite{pythia8,Sjostrand:2014zea} for the
fragmentation and hadronization of partons.

Signal samples for the $\PH\to\gamma^*\gamma\to\mu\mu\gamma$ produced
via $\Pg\Pg\PH$, VBF, and $\mathrm{V}\PH$ processes are simulated at NLO with
\MGvATNLO 2.3.3, with the Higgs boson characterization
framework~\cite{Artoisenet:2013puc,deAquino:2013uba}. The $\ttbar\PH$
production mechanism gives a negligible contribution to the signal and
is therefore ignored. For the $\PH\to\cPZ\gamma\to\ell\ell\gamma$
process, the simulated events from all four production mechanisms are
generated at NLO using \POWHEG v2.0~\cite{cite:powheg1,cite:powheg2}.
All signal samples are interfaced with {\PYTHIA\,8.212} with the
CUETP8M1 underlying event tune for hadronization and fragmentation.
The NLO PDF set, NNPDF3.0, is used to produce these samples. The SM
Higgs boson production cross sections and branching fractions
recommended by the LHC Higgs cross section working
group~\cite{LHC-YR4} are used for $\PH\to\cPZ\gamma$, whereas for
$\PH\to\gamma^{*}\gamma$ the Higgs boson production cross sections
are also taken from Ref.~\cite{LHC-YR4}, but the branching fraction of
$\PH\to\gamma^{*}\gamma$ is taken from the \MCFM calculation and given in
Eq.(~\ref{eq:fraction}).

The simulated signal and background events are reweighted by taking
into account the difference between data and simulated events so that
the distribution of pileup vertices, the trigger efficiencies, the
resolution, the energy scale, the reconstruction efficiencies, and the
isolation efficiency---for electrons, muons, and photons---observed
in data are reproduced. An additional correction is applied to photons
to reproduce the performance of the $\RNINE$ shower shape variable.

 The dominant backgrounds to $\PH\to\ell\ell\gamma$ consist of the
irreducible non-resonant SM $\ell\ell\gamma$ production, final-state
radiation in $\cPZ$ decays, $\gamma^*$ conversions, and Drell--Yan
production in association with jets, where a jet or a lepton is
misidentified as a photon.

The background is estimated from data, by fitting the observed
$\ell\ell\gamma$ mass distributions. Separate fits are performed to
the four event classes for the $\PH\to\gamma^*\gamma\to\mu\mu\gamma$
analysis and the thirteen classes for the
$\PH\to\cPZ\gamma\to\ell\ell\gamma$ analysis.  For the
$\PH\to\gamma^*\gamma\to\mu\mu\gamma$
($\PH\to\cPZ\gamma\to\ell\ell\gamma$) analysis, the range $110 (115)
< m_{\ell\ell\gamma} < 170\GeV$ is used in the fit.
 The fit model of the signal is
obtained from an unbinned fit to the mass distribution of the
corresponding sample of simulated events, using a double Crystal Ball
function~\cite{CB-Oreglia} in the
$\PH\to\gamma^*\gamma\to\mu\mu\gamma$ analysis, and a Crystal Ball
function plus a Gaussian function in the
$\PH\to\cPZ\gamma\to\ell\ell\gamma$ analysis.  To derive the signal
shapes for the intermediate mass points where simulation was not
available, a linear interpolation of the fitted parameters for
available mass points was performed.

The choice of the background fit function is based on a study that
minimizes  the bias that could be introduced by the selected
function. The study of the bias is performed for four families of
functions:

\begin{enumerate}
\item A sum of $N$ exponential functions
\begin{equation}
\sum\limits_{i=1}^{N}f_i\re^{p_i\,\mzg}
\end{equation}
with $2N$ free parameters: $p_i < 0$ and $f_i$.
The lowest order considered has $N=$ 1.

\item A sum of ${N}$ power-functions
\begin{equation}
\sum\limits_{i=1}^{N}f_i\mzg^{p_i} \,
\end{equation}
with $2{N}$ free parameters $p_i < 0$ and $f_i$.
The lowest order considered has $N=1$.

\item Bernstein polynomials of $N$th order, with $N =  1$, 2, 3, 4, and 5
\begin{equation}
\mathrm{Ber}_N(\mzg) = \sum\limits_{i=1}^{N}f_{i}^{2} \binom N i \mzg^{i}(1-\mzg)^{N-i}
\end{equation}
with $N$ free parameters $f_{i}$.

\item Laurent series with $N =  2$, 3, and 4 terms
\begin{equation}
f_2\mzg^{-4}+f_3\mzg^{-5},
\end{equation}
\begin{equation}
f_1\mzg^{-3}+f_2\mzg^{-4}+f_3\mzg^{-5},
\end{equation}
and
\begin{equation}
f_1\mzg^{-3}+f_2\mzg^{-4}+f_3\mzg^{-5}+f_4\mzg^{-6},
\end{equation}
with $N$ free parameters $f_{1\cdots N}$.
\end{enumerate}

A test is then performed to determine the best order in each
family. This test uses the difference in the negative log-likelihood
($NLL$) between the fits performed to data with two different orders of
 the same family of functions. 
The test starts with the lowest order $N$ in that family of functions
and the order is increased to the $(N{+}M)^{th}$ order until  
 the data support the hypothesis of the higher-order function.
For this purpose, a $p$-value of this quantity is calculated as:
\begin{equation}
p\text{-value} = \mathrm{Prob}(2\Delta{NLL} > 2\Delta{NLL}_{N+M} | \chi^{2}(M)),
\end{equation}
where $\Delta NLL$ is the difference of log-likelihood between the two
fits; $\Delta NLL_{N+M} = 2(NLL_{N} -
NLL_{N+M})$ follows a $\chi^{2}$ distribution with $M$
degrees of freedom, where $M$ is the difference in the number of
free parameters between the $N{+}M$ function and the $N$
function; $NLL_{N}$ and $NLL_{N+M}$ are the values of
the log-likelihood of the fit to data using $N^{th}$ and
$(N{+}M)^{th}$ order functions from a family.  If
the $p$-value is less than 0.05, the higher order function is
supported by the data and the procedure is then applied to other
higher order functions in the same family.  The procedure stops when
the $p$-value becomes greater than 0.05.

Once the best order of each family is determined for each category,
pseudo-experiments (with no injected signal) describing possible
experimental outcomes are randomly generated using each of the
determined functions as generators of background.  A
signal-plus-background fit is performed for each of these sets of
pseudo-experiments with all other background functions of the chosen
order, so that the presence of a possible bias introduced by the
fitting function can be determined. In each fit, the bias is estimated
with a pull variable, computed as ($\mu_\mathrm{FIT}
- \mu_\mathrm{t})/\sigma_\mathrm{FIT}$, where $\mu_\mathrm{FIT}$ and
$\sigma_\mathrm{FIT}$ are the mean and the standard deviation of the
signal strength determined from the signal-plus-background fit, and
$\mu_\mathrm{t}$ is the true injected signal strength, which is zero in this
case.  A given fit function is deemed acceptable in a given category
if its pull is less than 0.14 when
fitting pseudo-experiments generated with all of the other functional
families. With this requirement, the error on the frequentist coverage
of the quoted measurement in the analysis is less than 1\%, where
the coverage is defined as the fraction of experiments in which the
true value is contained within the confidence
interval. If several functions pass this criterion,
 then we choose the one which has the least pull. 
Table~\ref{tab:bkgfits} shows the fit functions chosen in
each category of the analysis.

\begin{table*}[hbt]
\centering
\topcaption{Fit functions chosen as a result of the bias study used in the analysis.
\label{tab:bkgfits}
}
\begin{tabular}{ccc}
$m_{\ell\ell}$ & Category & Best fit function \\
\hline
 \multirow{4}{*}{$<$50\GeV}
& EB, high $\RNINE$  & Bernstein of order 4\\
& EB, low $\RNINE$  & Bernstein of order 4\\
& EE   & Bernstein of order 4\\
& Dijet tag & Exponential of order 2\\[\cmsTabSkip]
\multirow{7}{*}{$>$50\GeV}
& Lepton tag & Power law of order 1\\
& Dijet tag & Power law of order 1\\
& Boosted & Bernstein of order 3\\
& Untagged 1 & Bernstein of order 4\\
& Untagged 2 &  Bernstein of order 5\\
& Untagged 3 &  Bernstein of order 4\\
& Untagged 4 &  Bernstein of order 4\\
\end{tabular}
\end{table*}

\begin{figure*}[hbtp]
  \centering
    \includegraphics[width=1.2\cmsFigWidth]{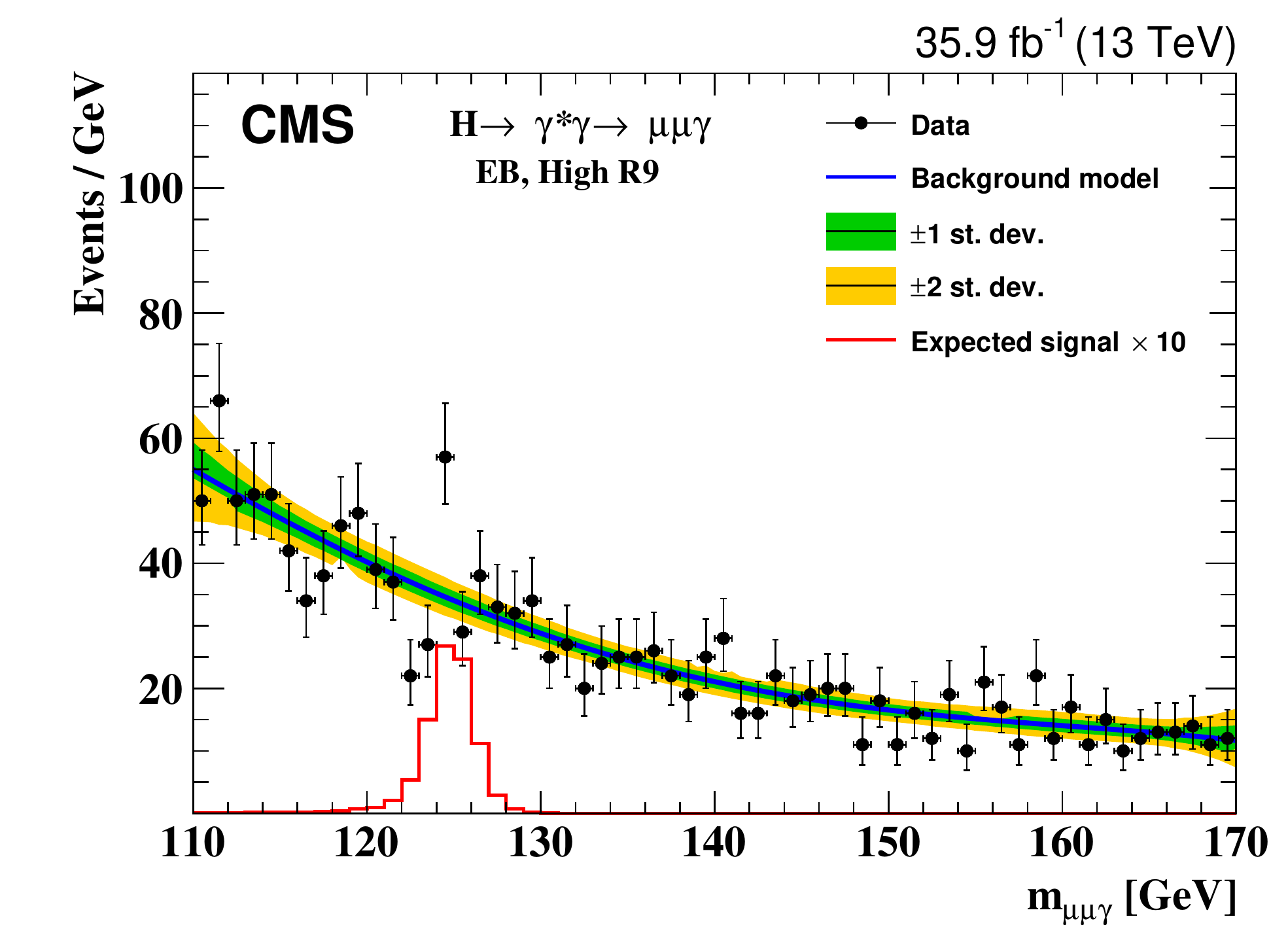}
    \includegraphics[width=1.2\cmsFigWidth]{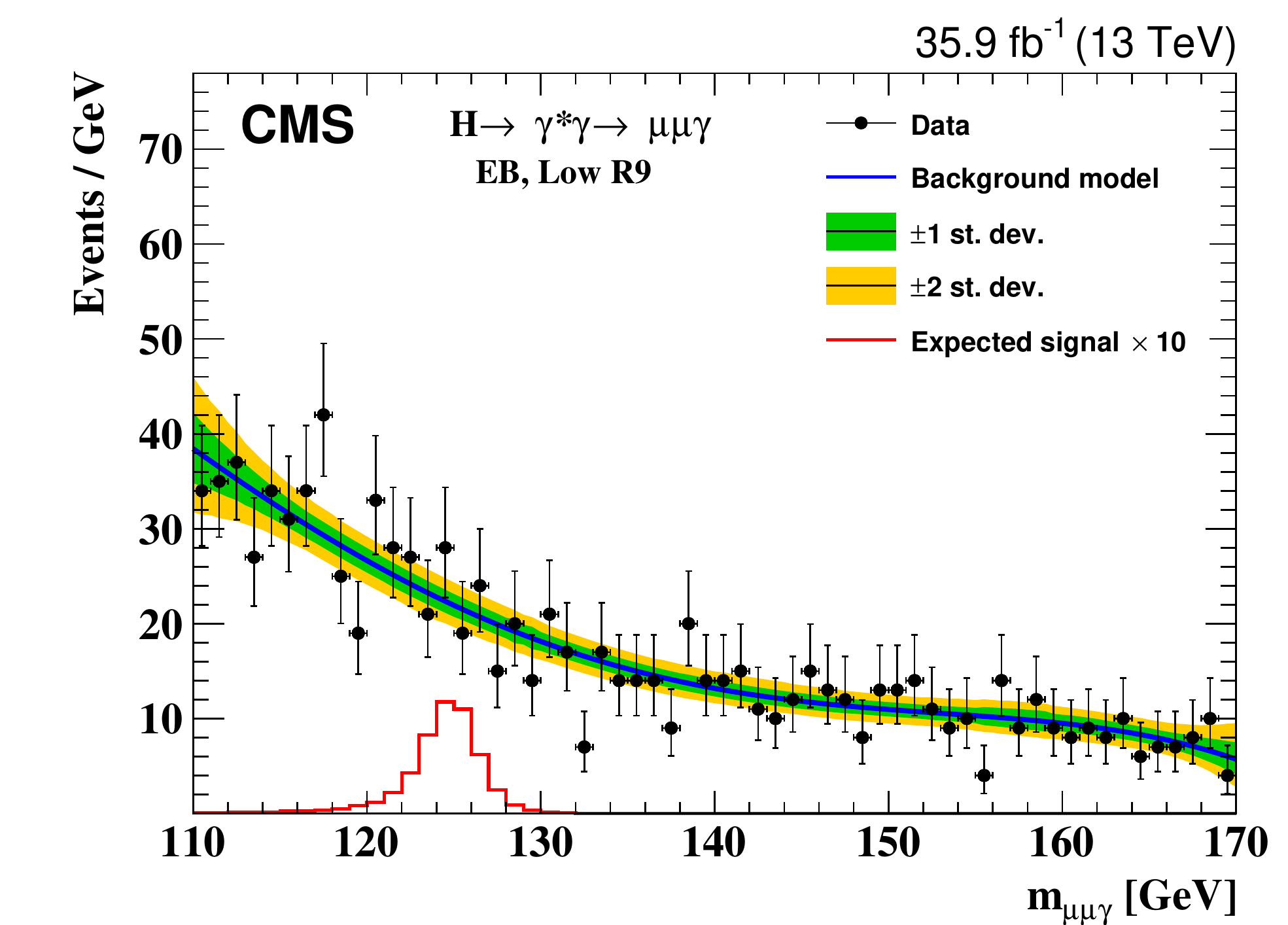}\\
    \includegraphics[width=1.2\cmsFigWidth]{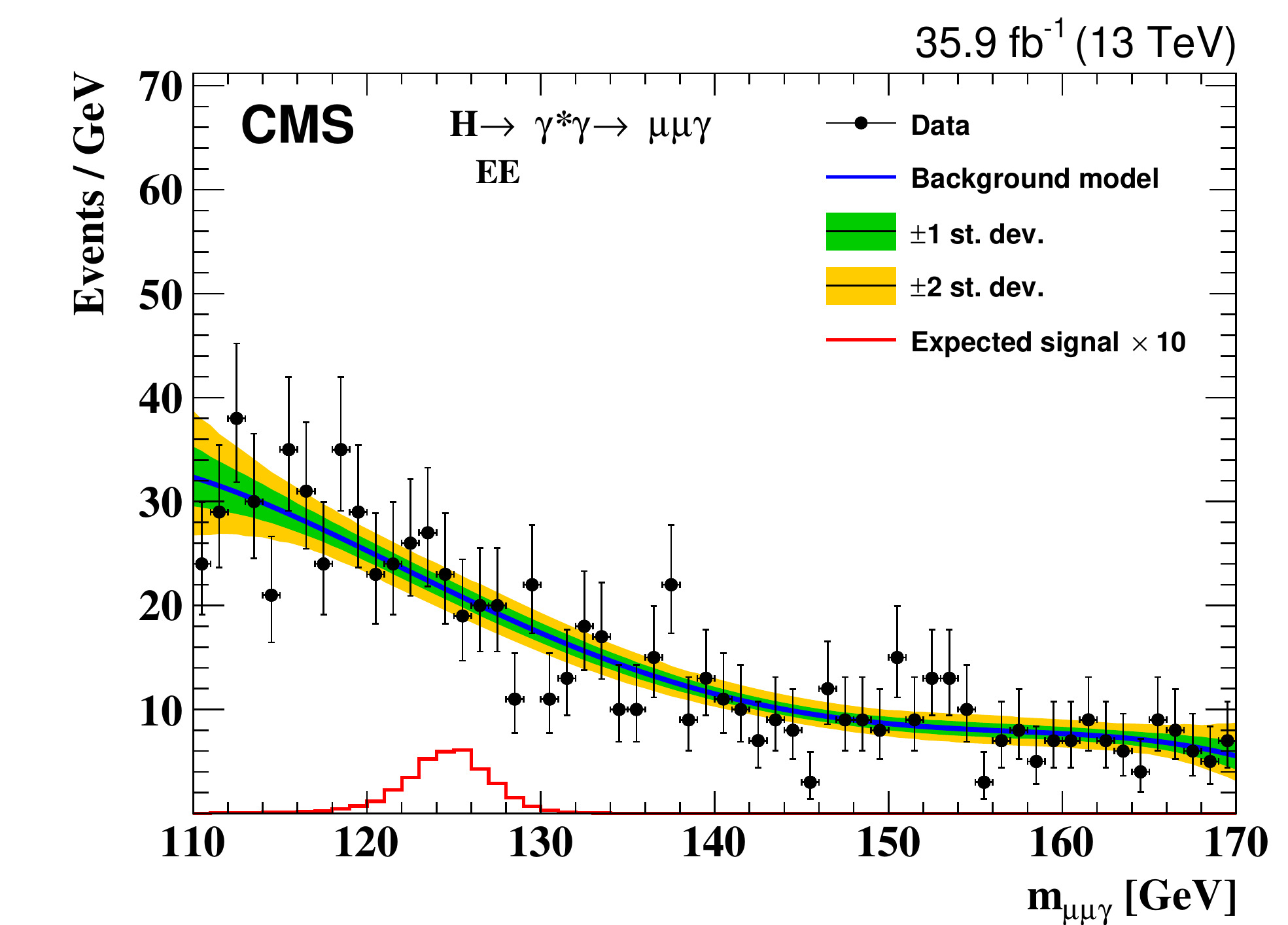}
    \includegraphics[width=1.2\cmsFigWidth]{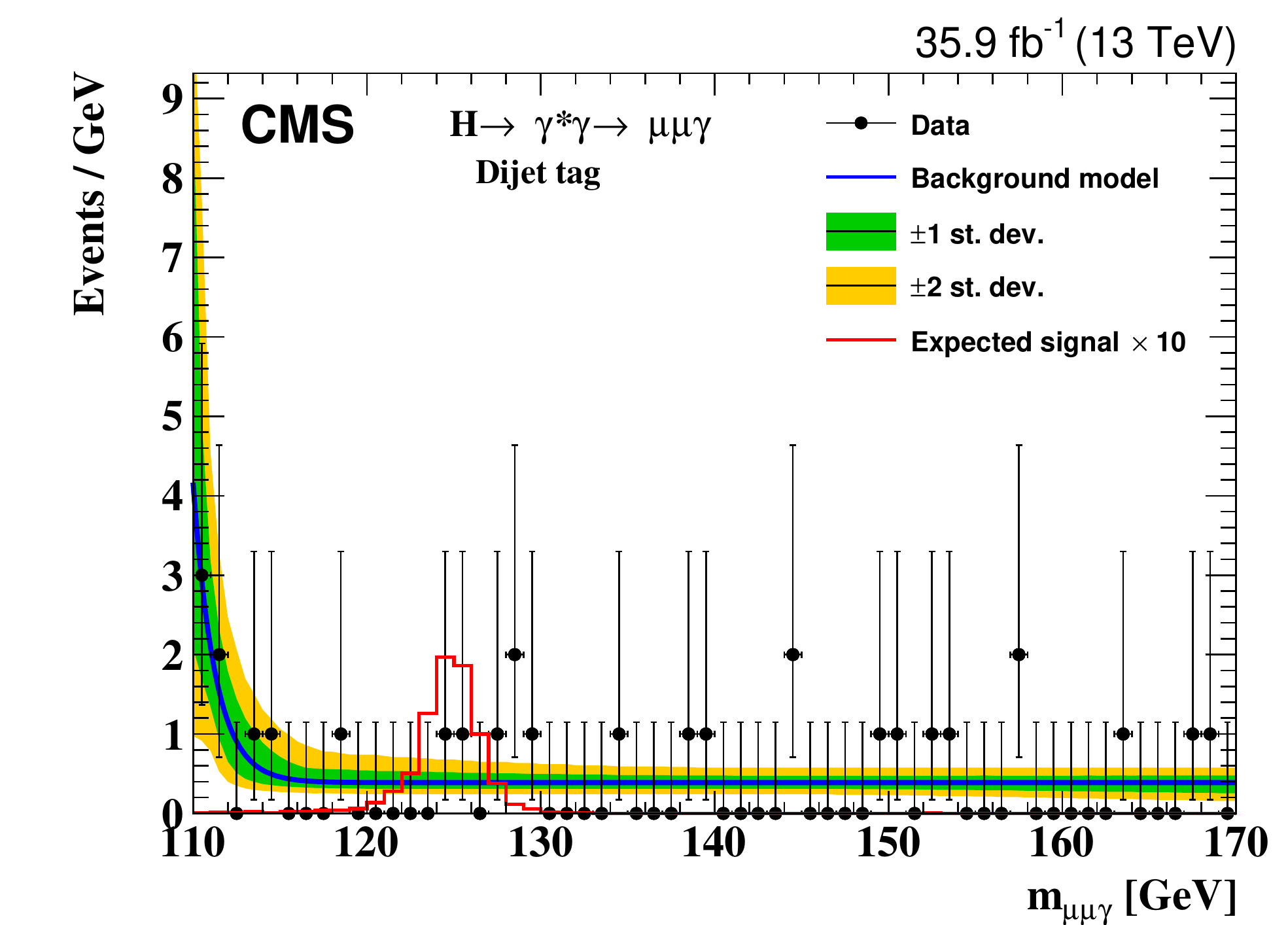}
    \caption{
        Background model fit to the $m_{\mu\mu\gamma}$ distribution for
EB-high $\RNINE$ (upper left), EB-low $\RNINE$ (upper right), EE
(lower left) and dijet tag (lower right) for the
$\PH\to\gamma^*\gamma\to\mu\mu\gamma$ selection.  The green and yellow
bands represent the 68 and 95\% \CL uncertainties in the fit to the
data.  \label{fig:3D}}
\end{figure*}

\begin{figure*}[htb]
\centering
\includegraphics[width=0.4\textwidth]{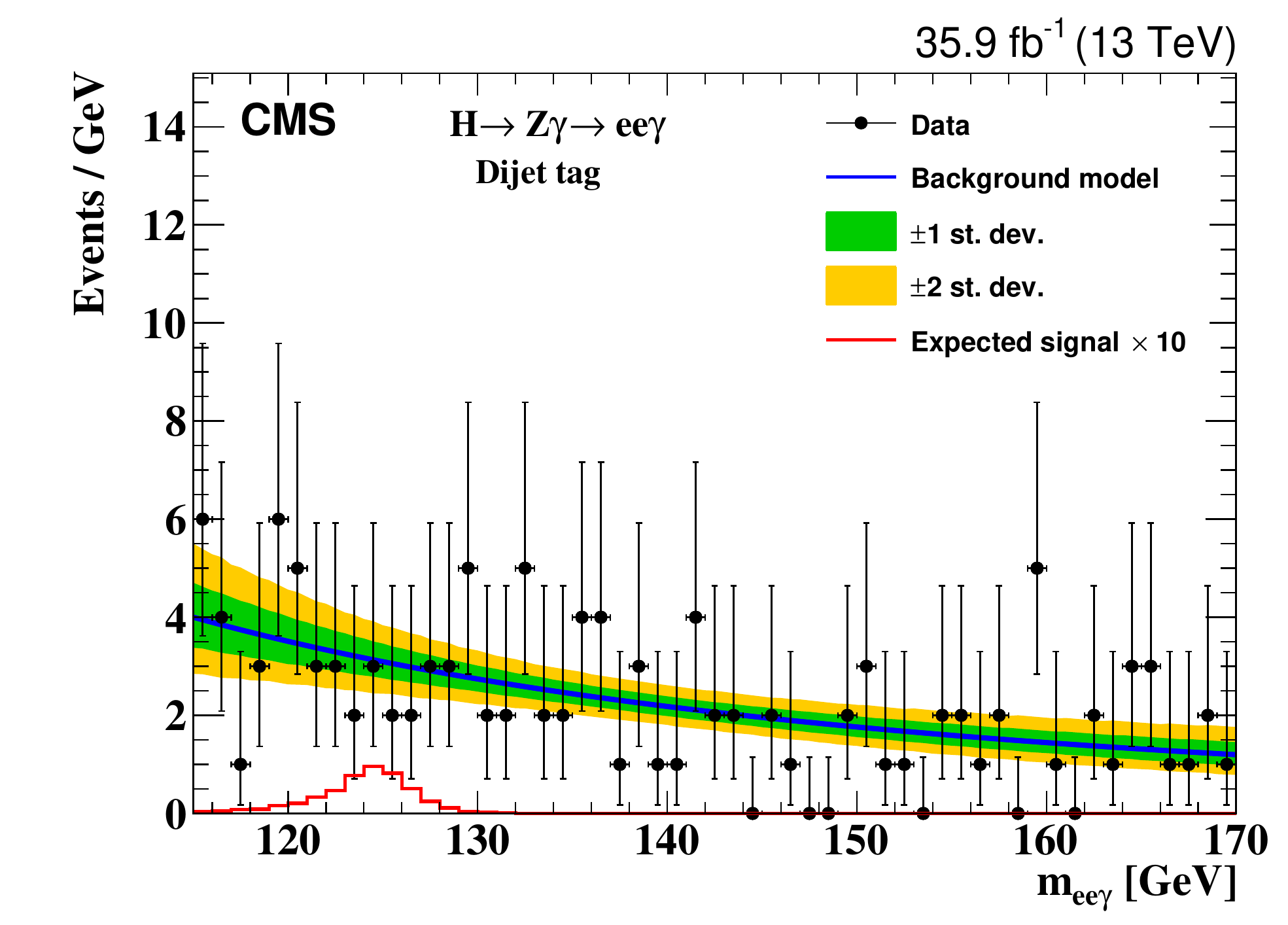}
\includegraphics[width=0.4\textwidth]{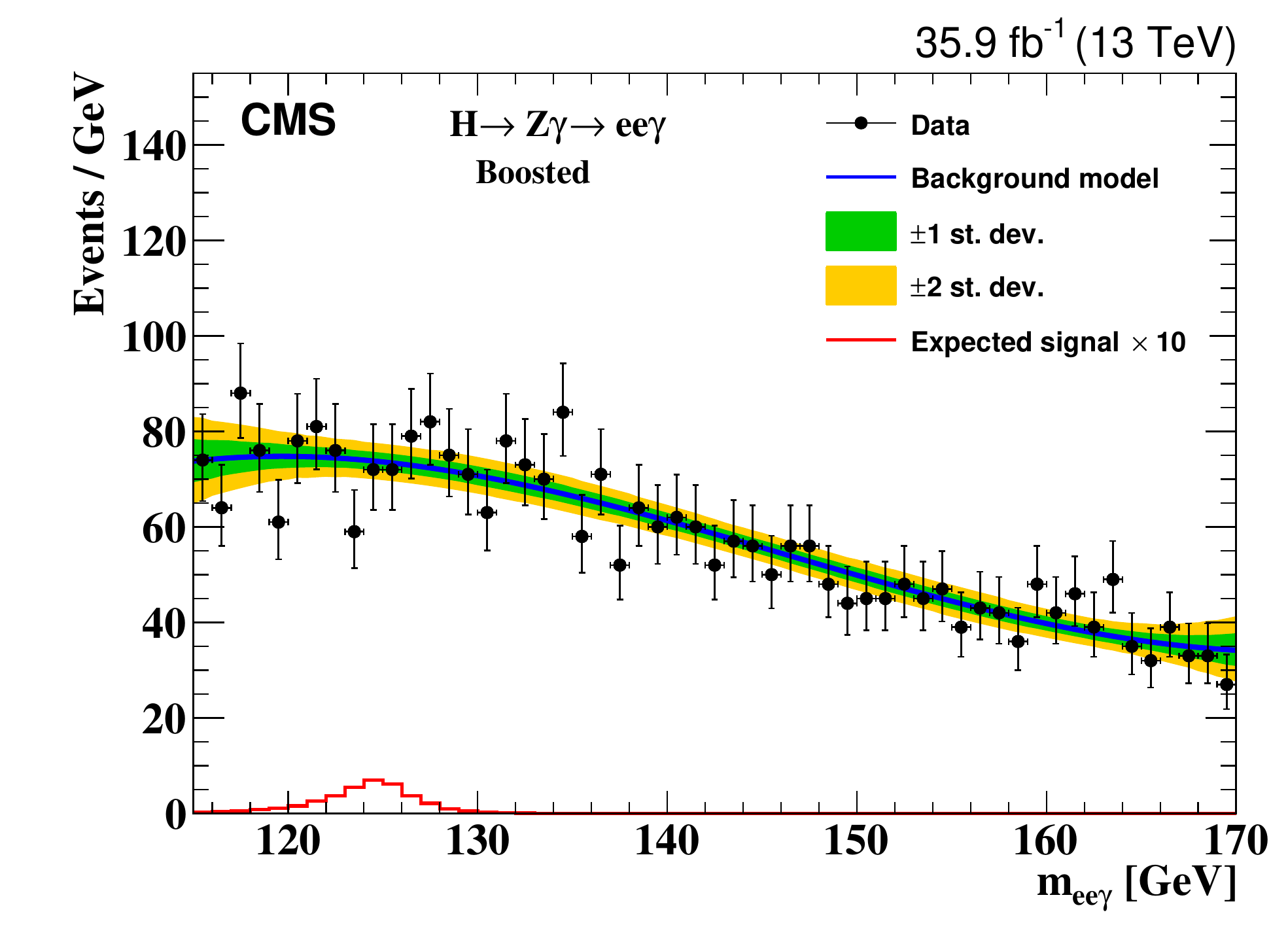}\\
\includegraphics[width=0.4\textwidth]{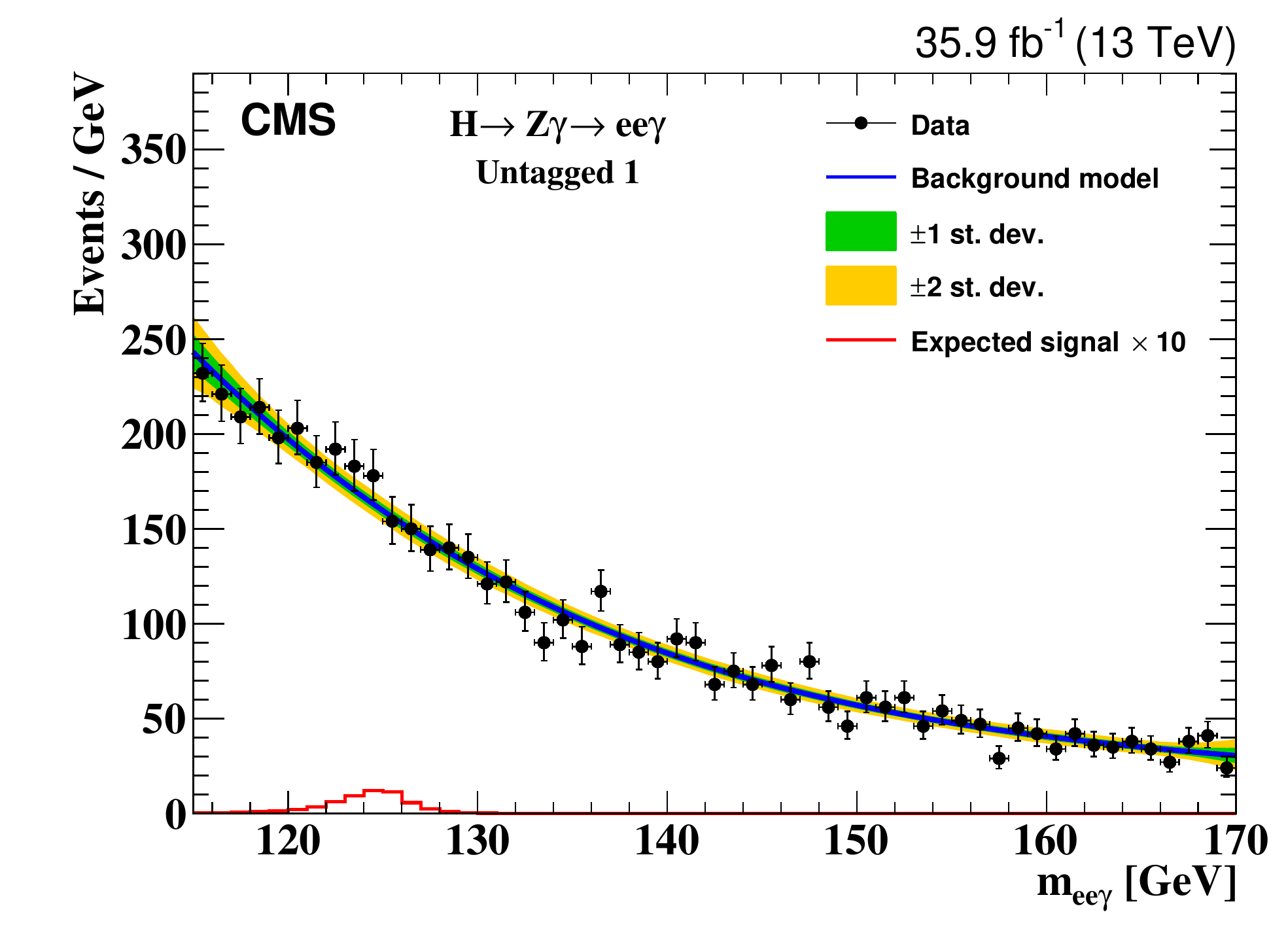}
\includegraphics[width=0.4\textwidth]{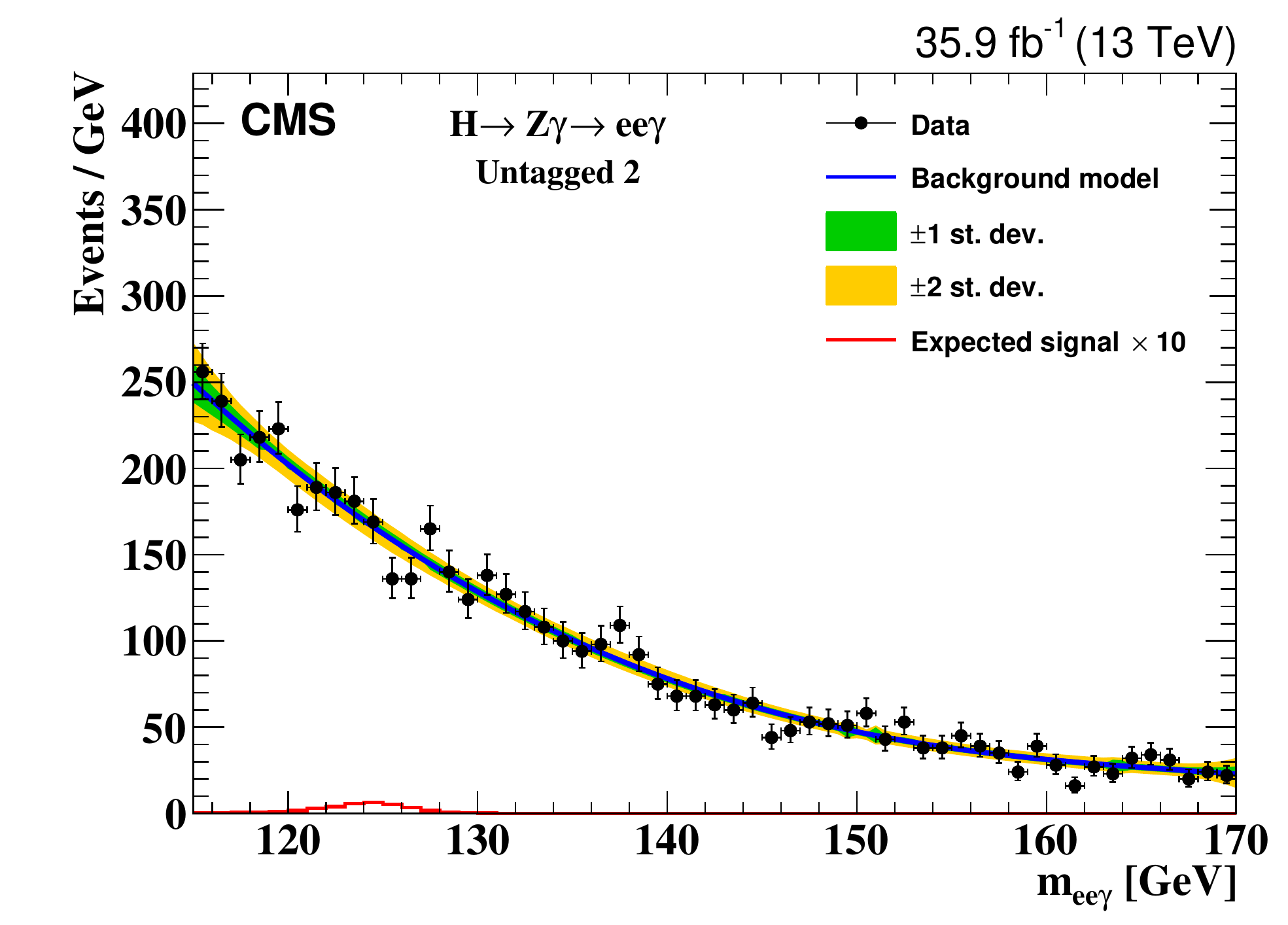}\\
\includegraphics[width=0.4\textwidth]{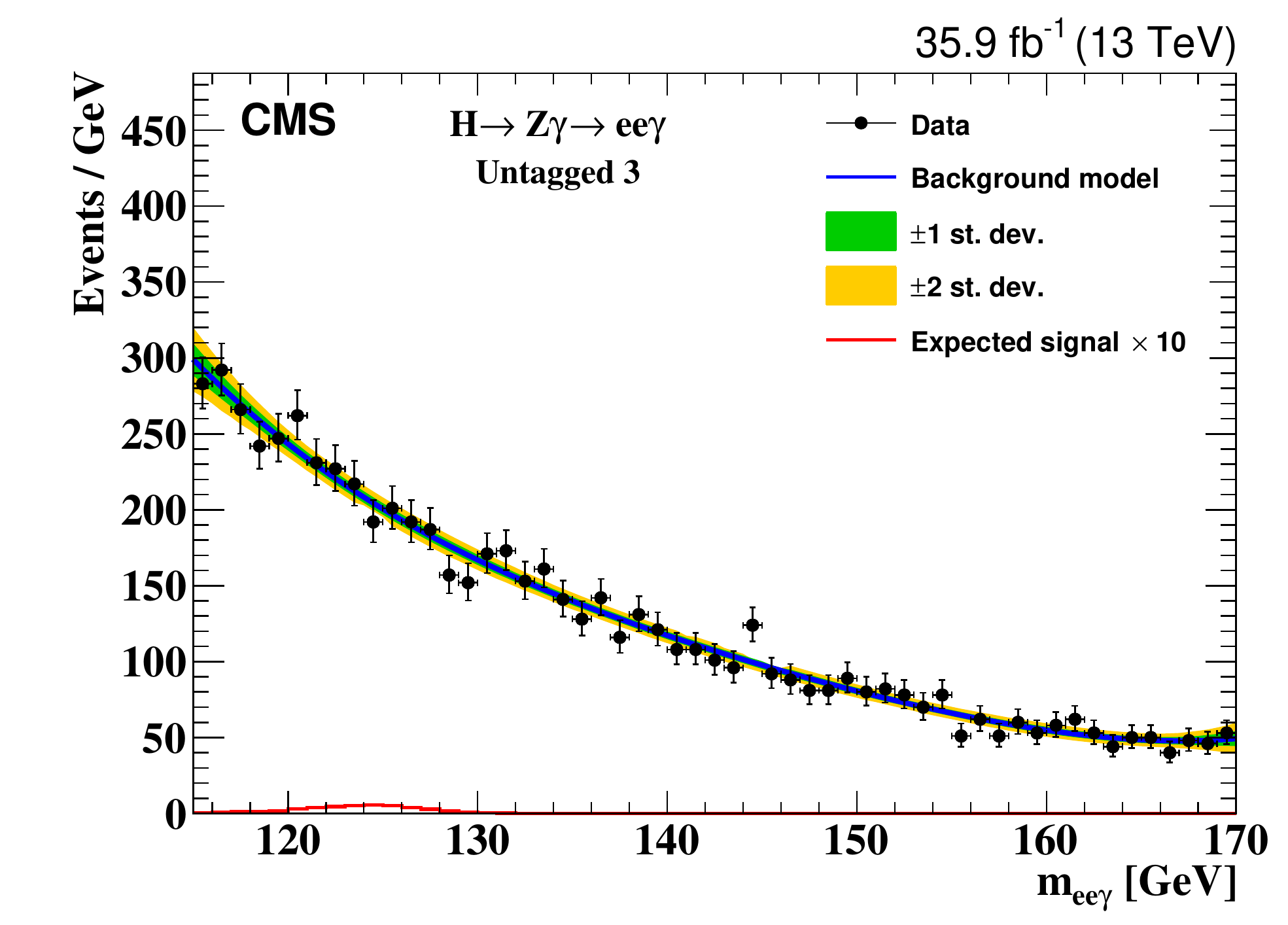}
\includegraphics[width=0.4\textwidth]{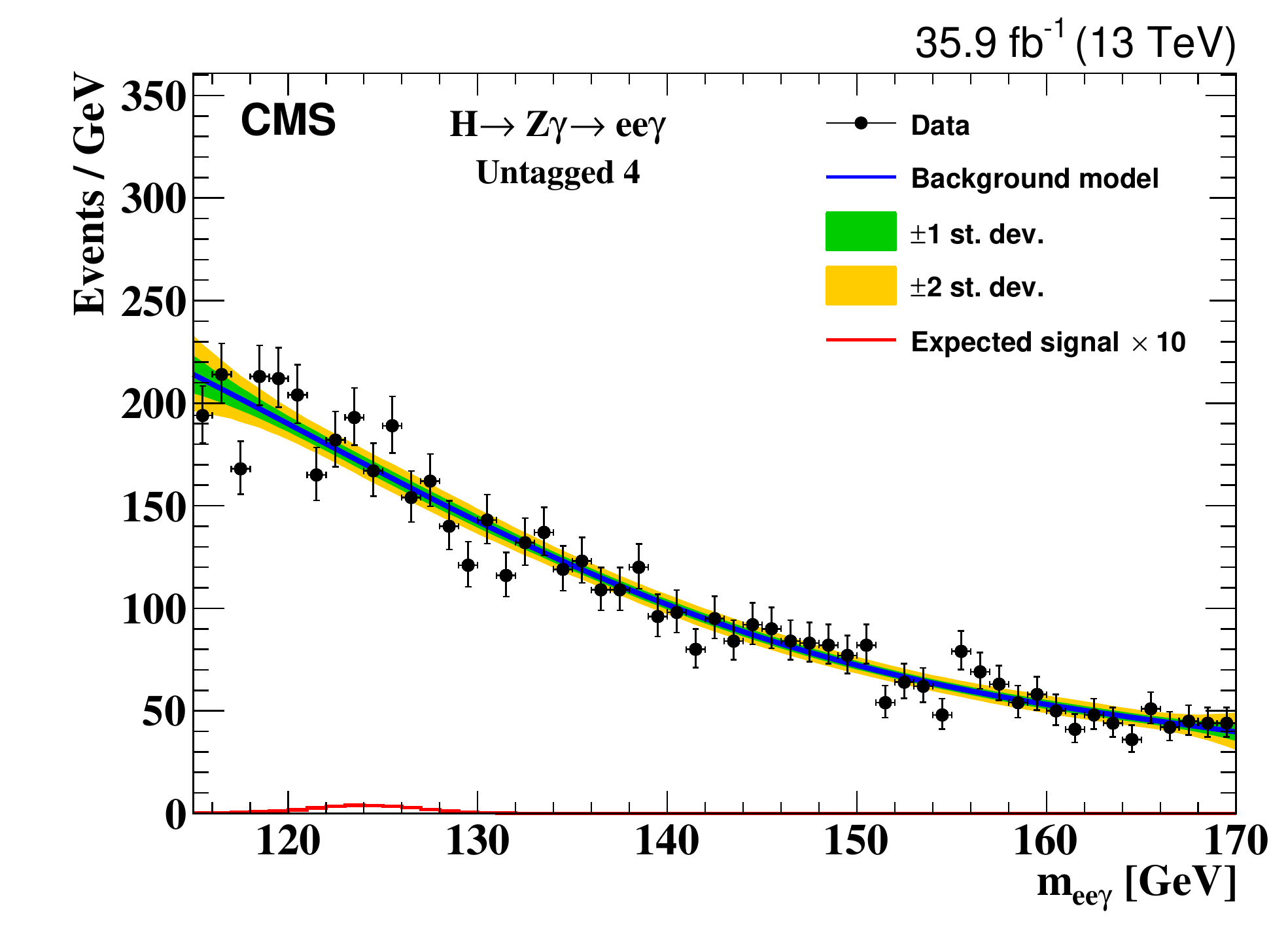}\\
 \caption{Background model fit to the $m_{\Pe\Pe\gamma}$ distribution
for dijet tag (upper left), boosted (upper right), untagged 1 (middle
left), untagged 2 (middle right), untagged 3 (bottom left), and
untagged 4 (bottom right) for the $\PH\to\cPZ\gamma\to\Pe\Pe\gamma$
selection.  The green and yellow bands represent the 68 and 95\% \CL
uncertainties in the fit to the data.  \label{fig:3el}}
\end{figure*}

\begin{figure*}[htb]
\centering
\includegraphics[width=0.4\textwidth]{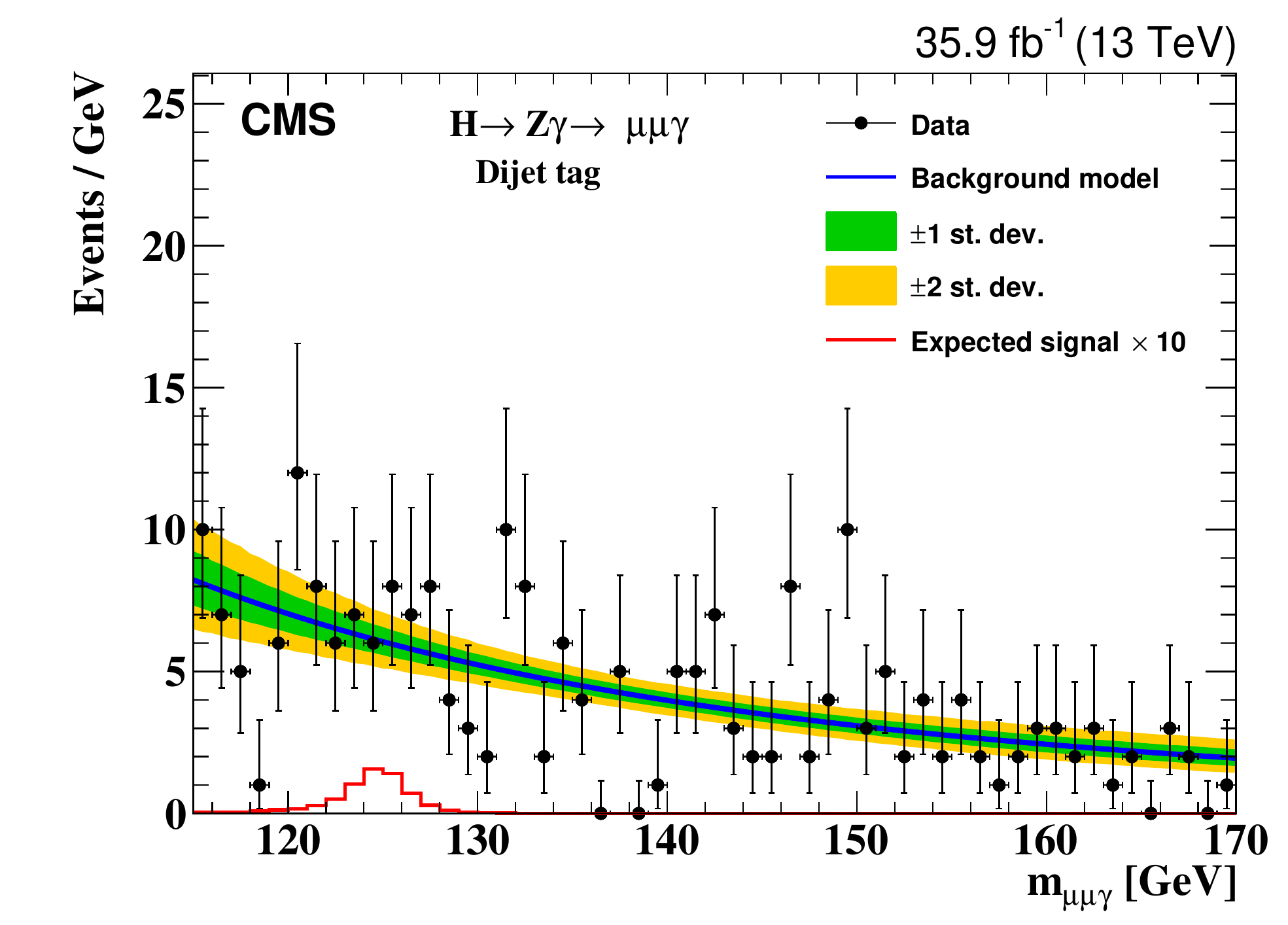}
\includegraphics[width=0.4\textwidth]{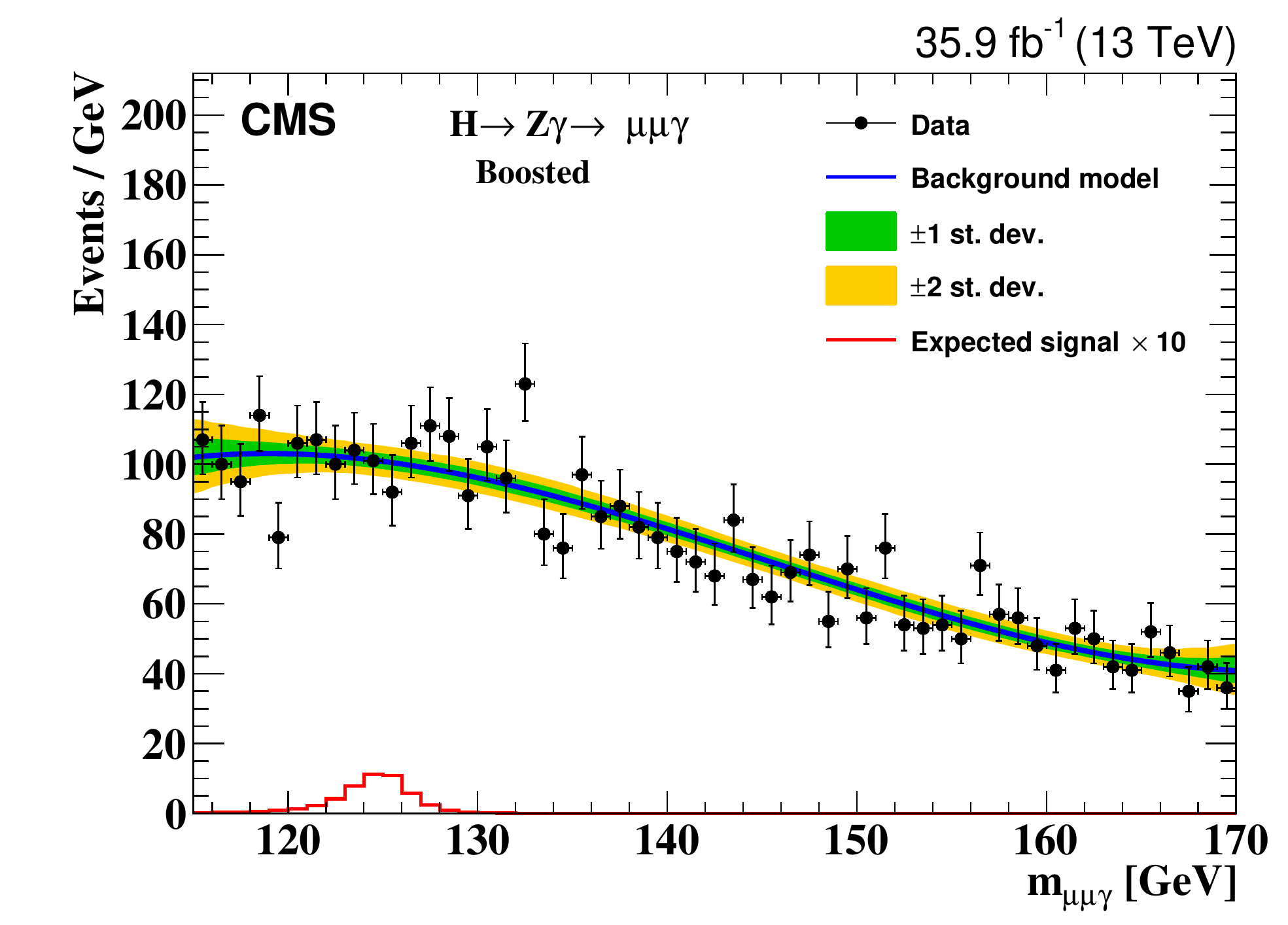}\\
\includegraphics[width=0.4\textwidth]{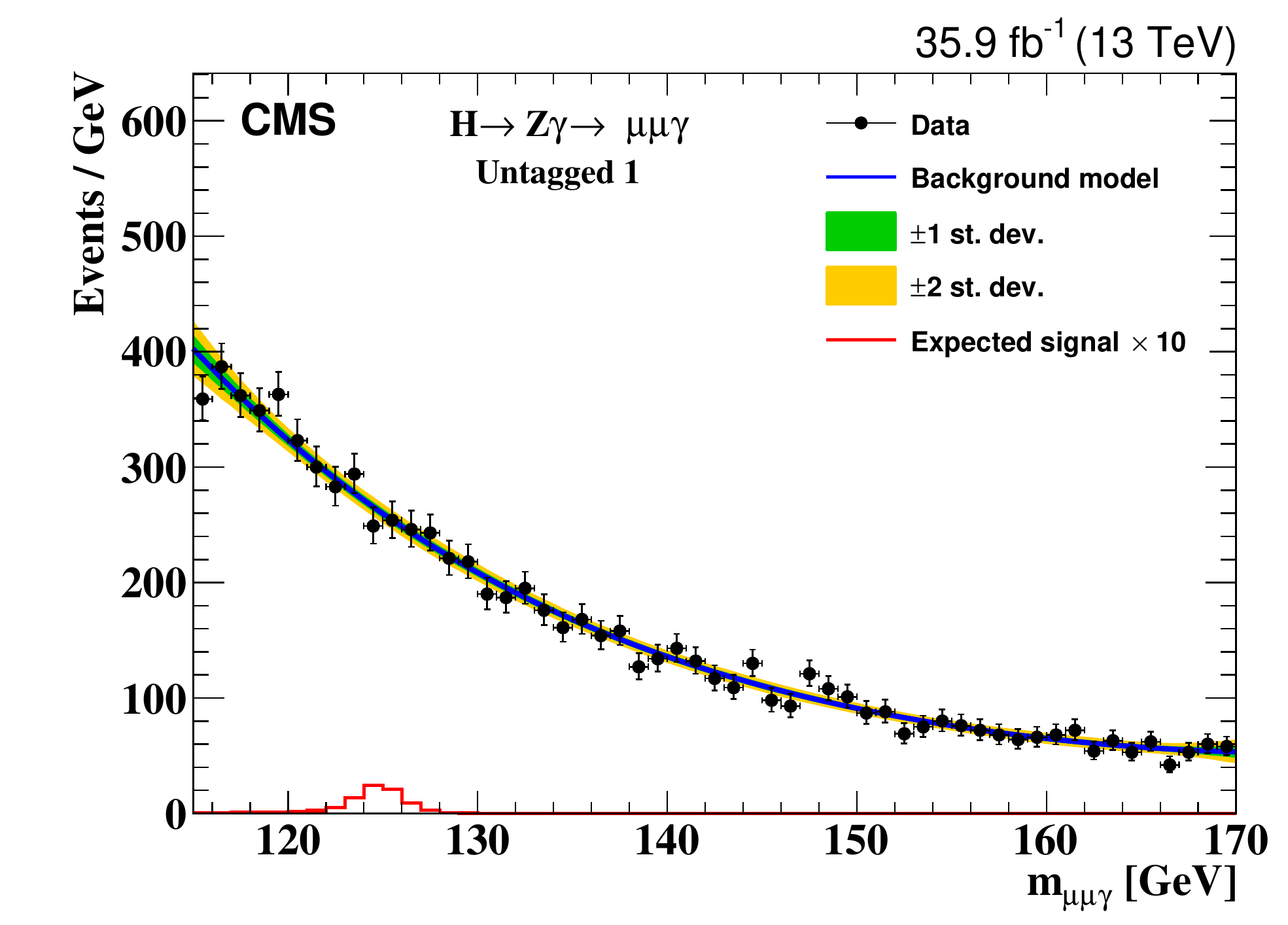}
\includegraphics[width=0.4\textwidth]{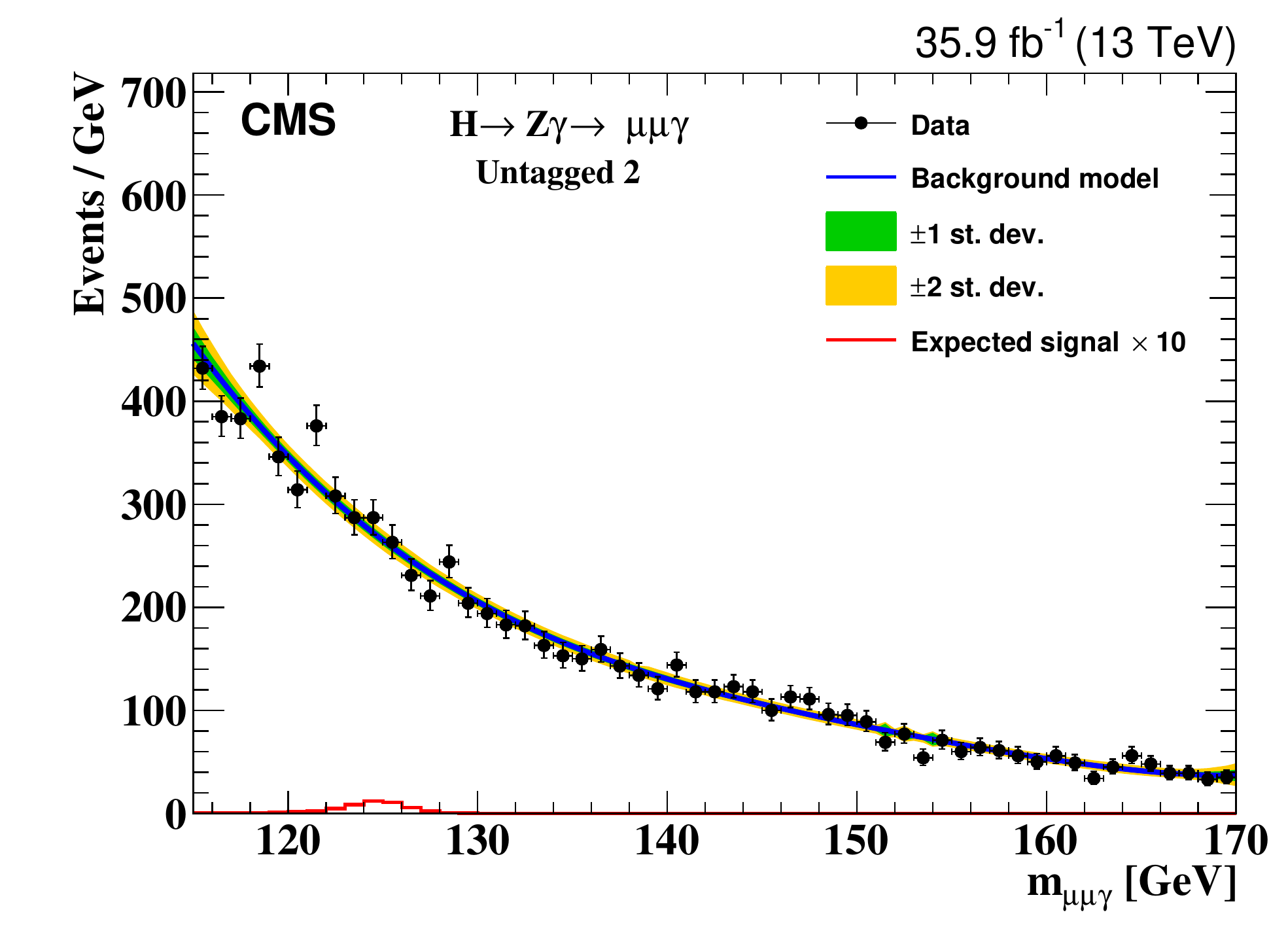}\\
\includegraphics[width=0.4\textwidth]{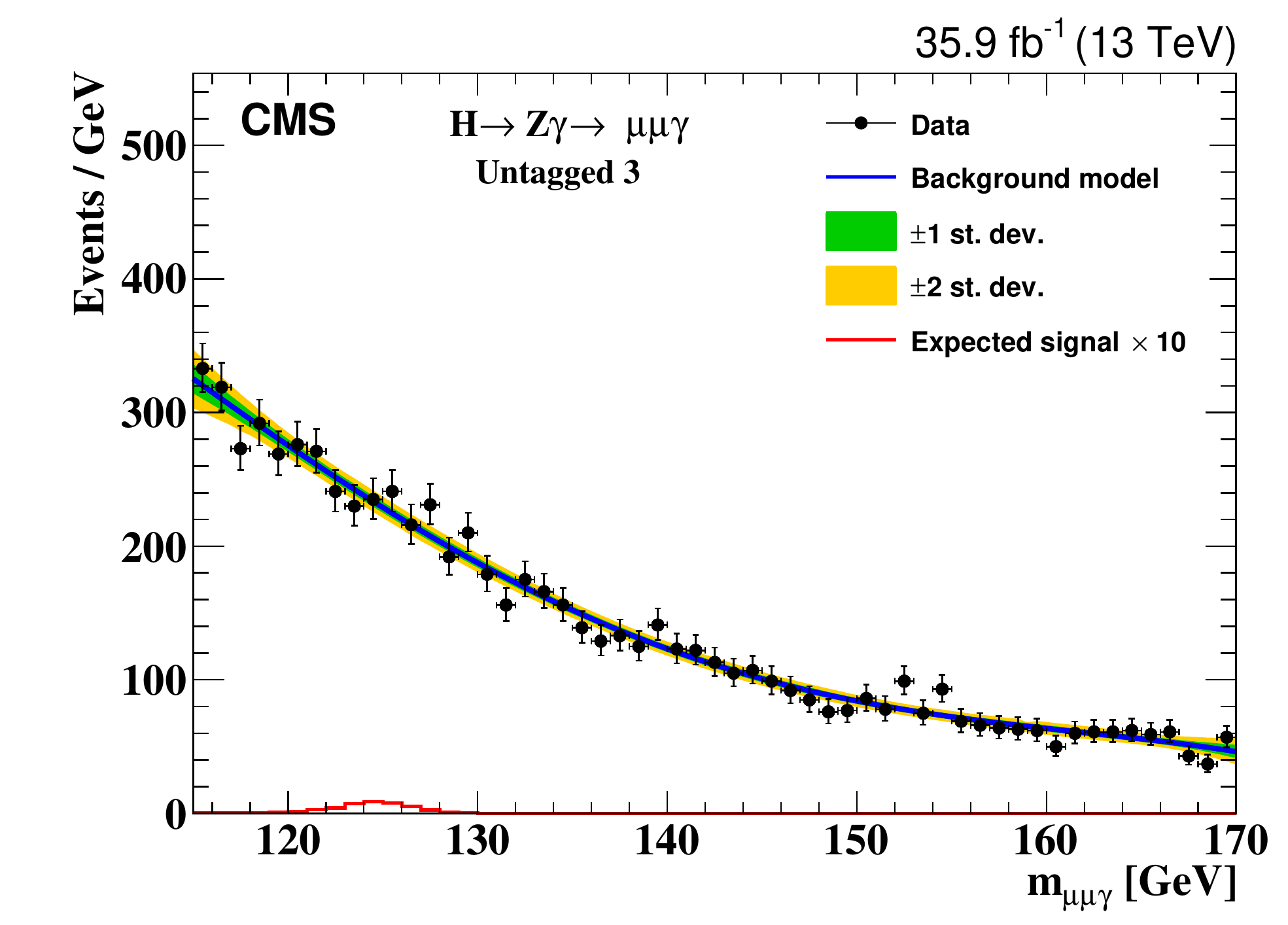}
\includegraphics[width=0.4\textwidth]{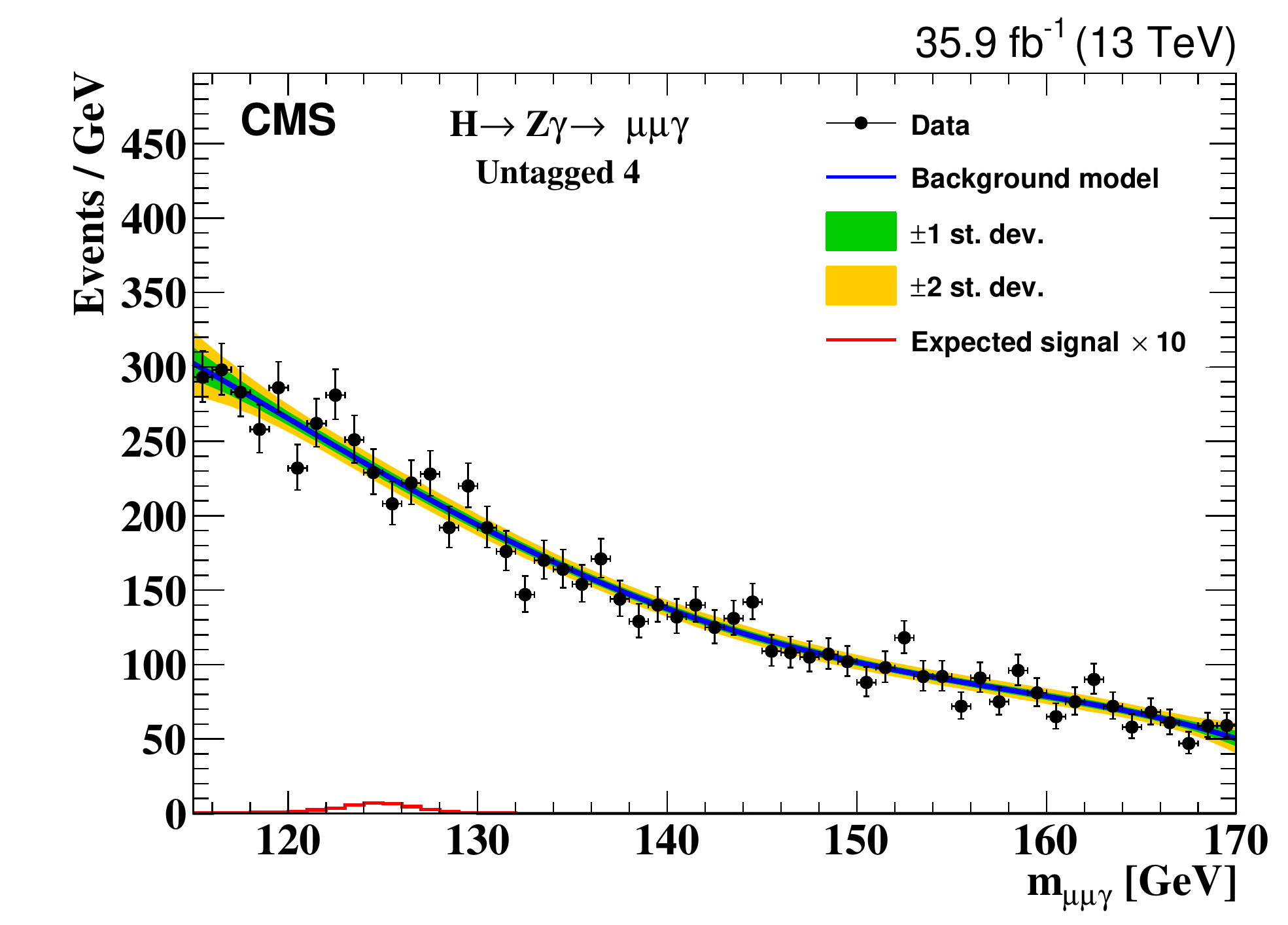}\\
 \caption{Background model fit to the $m_{\mu\mu\gamma}$ distribution
for dijet tag (upper left), boosted (upper right), untagged 1 (middle
left), untagged 2 (middle right), untagged 3 (bottom left), and
untagged 4 (bottom right) for the $\PH\to\cPZ\gamma\to\mu\mu\gamma$
selection.  The green and yellow bands represent the 68 and 95\% \CL
uncertainties in the fit to the data.  \label{fig:3mu}}
\end{figure*}

\begin{figure}[htb]
\centering
\includegraphics[width=0.4\textwidth]{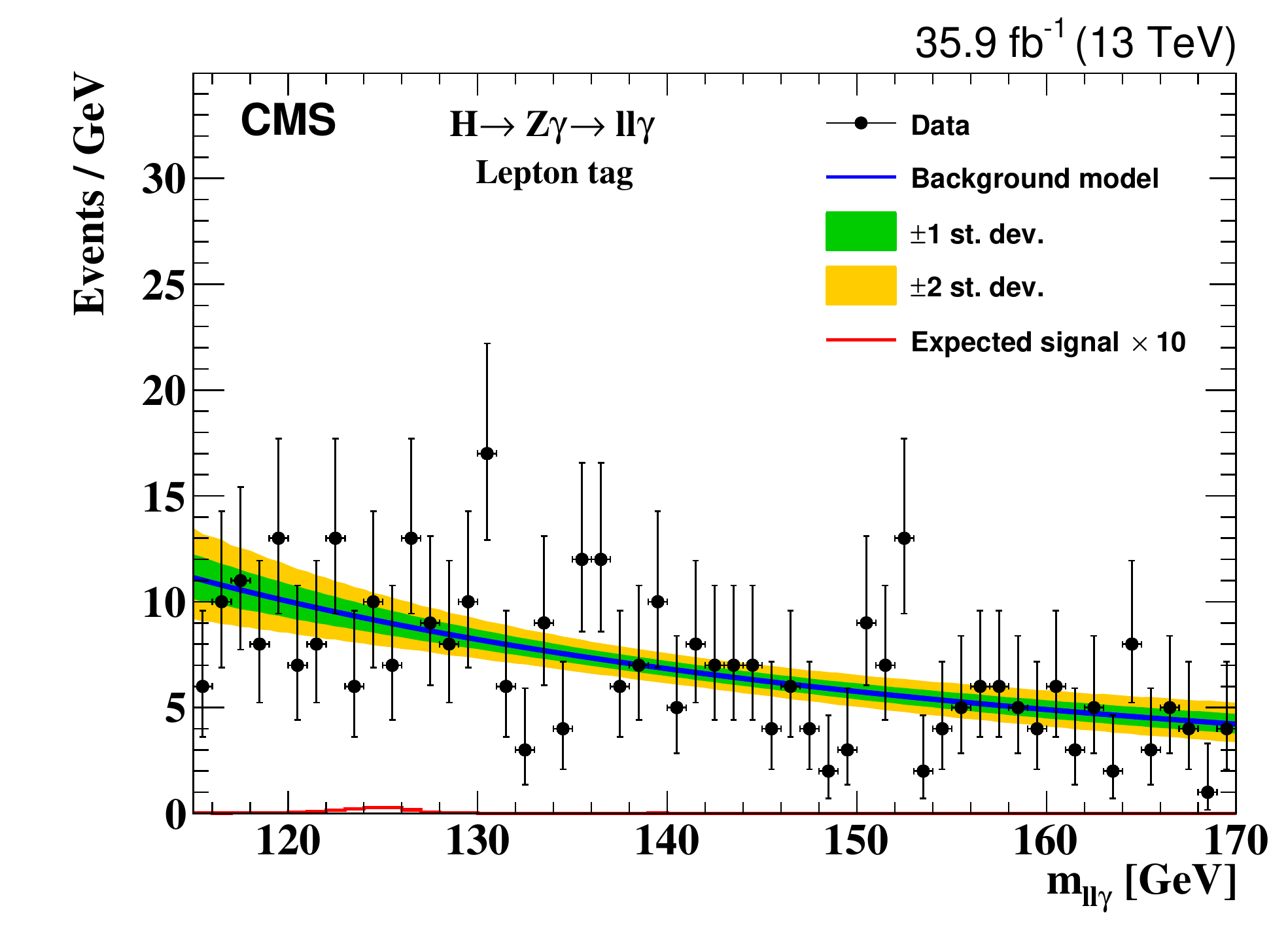}
 \caption{Background model fit to the $m_{\ell\ell\gamma}$
distribution for $\PH\to\cPZ\gamma\to\ell\ell\gamma$ lepton tag
category.  The green and yellow bands represent the 68 and 95\% \CL
uncertainties in the fit to the data.  \label{fig:bkg_leptag}}
\end{figure}

The background fits based on the $m_{\ell\ell\gamma}$ data
distributions for the event categories of the
$\PH\to\gamma^*\gamma\to\mu\mu\gamma$ analysis are shown in
Fig.~\ref{fig:3D} and, for the electron and muon channels in all
$\PH\to\cPZ\gamma\to\ell\ell\gamma$ event class definitions except 
for the lepton tag category, in Figs.~\ref{fig:3el} and \ref{fig:3mu} respectively. 
Finally, Fig.~\ref{fig:bkg_leptag} shows the background fit for the lepton tag
category in the $\PH\to\cPZ\gamma\to\ell\ell\gamma$ analysis. 
As we can see from these figures, the background fits describe the data
well.

\clearpage

\section{Systematic uncertainties and results}
\label{sec:results}

No significant deviation from the background-only hypothesis is
observed. The data are used to derive upper limits on the Higgs boson
production cross section times the branching fractions, $\sigma(\Pp\Pp\to\PH)\,\mathcal{B}(\PH\to\gamma^*\gamma\to\mu\mu\gamma)$ and
$\sigma(\Pp\Pp\to\PH)\,\mathcal{B}(\PH\to\cPZ\gamma\to\ell\ell\gamma)$, divided
by the corresponding SM predictions. The limits are evaluated using a
modified frequentist approach, asymptotic \CLs, taking the profile
likelihood as a test statistic~\cite{cite:l1,cite:l2,cite:l3,
Cowan:2010js}. An unbinned evaluation of the likelihood is considered.

Background uncertainties are taken from the fit to the data. 
The sources of systematic uncertainties related to the signal are listed below.
 The first two sources affect the signal shape and the remaining sources affect the signal yield.

\begin{itemize}

\item Electron and photon energy scale and resolution: The electromagnetic energy scale
is known with 0.15--0.5 (1)\% precision in EB (EE). To quantify the
corresponding uncertainty, the electron and photon energies are varied
and the effects on signal mean and resolution are propagated as shape
nuisance parameters in the estimation of limits.

\item Muon momentum scale and resolution: The uncertainty in the muon momentum scale
is 1\%. To quantify the corresponding uncertainty, the muon momentum scale is
varied and the effect on signal mean and resolution is propagated as a
shape nuisance parameter in the estimation of limits.

\item Integrated luminosity: The uncertainty in the integrated luminosity is 2.5\%~\cite{LUM-17-001}.
  This is applied as a normalization uncertainty to the total expected
  yield of the signal.

\item Object identification and isolation: The corrections applied to the simulation
to reproduce the performance of the lepton and photon selection are
measured with $\cPZ\to\Pe\Pe$ and $\cPZ\to\mu\mu$
events.

\item Pileup: The uncertainty from the description of the pileup in the signal
simulation is estimated by varying the total inelastic cross section
by $\pm$4.6\% ~\cite{Sirunyan:2018nqx}.

\item Jet-energy scale and resolution: The uncertainties in the jet energy scale and
resolution are accounted for by changing the jet response and
resolution by $\sim$2$\%$.

\item Underlying event and parton shower uncertainty: The uncertainty associated with the choice and tuning of the
generator is estimated with dedicated samples which are generated by
varying the parameters of the tune used (CUETP8M1) to generate the
original signal samples.  The difference in signal yields with respect
to the nominal configuration is propagated as the uncertainty.

\item $\RNINE$ reweighting:  This shower-shape variable in the signal simulation is reweighted
to match that in the data.  This reweighting introduces an uncertainty
 that is estimated by removing the $\RNINE$ reweighting in the
 simulation and then estimating the yields in the categories where
 $\RNINE$ is used for categorization.

\item Theoretical uncertainties: These include the systematic uncertainties from the
  effect of the choice of PDF on the signal
  cross section~\cite{cite:cs1,cite:cs2,Butterworth:2015oua} and the
  uncertainty in the Higgs boson branching fraction
  prediction. The uncertainty in the branching ratio
  of $\PH\to\cPZ\gamma$ is calculated to be 5.6\%~\cite{LHC-YR4}. In the case of $\PH\to\gamma^*\gamma$
  analysis, there is no available theoretical uncertainty. 
  So it is taken by rounding off the error on the branching ratio of $\PH\to\cPZ\gamma$ to 6\%.

\end{itemize}

The pre-fit values of the nuisance parameters, averaged over all the
 categories, are summarized in Table~\ref{tab:syst}.

\begin{table*}[!htb]
\centering
\topcaption{Sources of systematic uncertainties considered in the
  $\PH\to\cPZ\gamma\to\ell\ell\gamma$ and
  $\PH\to\gamma^*\gamma\to\mu\mu\gamma$ analyses.  The pre-fit values
  of the nuisance parameters are shown averaged over all the
  categories in the analysis which either affect the normalization of
  the simulated signal event yields or the mean and resolution of
  $m_{\ell\ell\gamma}$. The ``\NA'' indicates that the uncertainty is
  not applicable.
  \label{tab:syst}}
{\small
\begin{tabular}{lcc}
Sources                 &  $\PH\to\cPZ\gamma\to\ell\ell\gamma$  & $\PH\to\gamma^*\gamma\to\mu\mu\gamma$ \\ \hline
{Theory} &    \\
-- $\Pg\Pg\PH$ cross section (scale) & $3.9\%$ & $3.9\%$\\
-- $\Pg\Pg\PH$ cross section (PDF)& $3.2\%$ & $3.2\%$\\
-- VBF cross section (scale)&  $+0.4\% -0.3\%$ & $+0.4\% -0.3\%$\\
-- VBF cross section (PDF)&  $2.1\%$ & $2.1\%$\\
-- $\PW\PH$ cross section (scale) & $+0.5\% -0.7\%$ & $+0.5\% -0.7\%$  \\
-- $\PW\PH$ cross section (PDF)& $1.9\%$ & $1.9\%$ \\\
-- $\cPZ\PH$ cross section (scale) & $+3.8\% -3.1\%$ &  $+3.8\% -3.1\%$ \\
-- $\cPZ\PH$ cross section (PDF)& $1.6\%$  &  $1.6\%$ \\
-- $\ttbar\PH$ cross section (scale)& $+5.8\% -9.2\%$ & \NA \\
-- $\ttbar\PH$ cross section (PDF)& $3.6\%$ & \NA \\
Underlying event and parton shower & & \\
-- Muon channel & $3\%$ & $4.7\%$\\
-- Electron channel & $3\%$& \NA \\
Branching fraction & $5.7\%$ & $6\%$\\
{Integrated luminosity}  & $2.5\%$  & $2.5\%$ \\
{Lepton identification and isolation}  &  & \\
-- Muon channel & $0.6\%$ & $2\%$\\
-- Electron channel & $1.2\%$ & \NA\\
{Photon identification and isolation}  &  & \\
-- Muon channel & $2.3\%$ & $1.6\%$\\
-- Electron channel & $2.2\%$ & \NA\\
{Pileup  reweighting} &  & \\
-- Muon channel & $0.6\%$ & $0.3\%$\\
-- Electron channel & $0.9\%$ & \NA\\
{\RNINE} {reweighting} &  & \\
-- Muon channel & $6.5\%$ & $9\%$ \\
-- Electron channel & $6.8\%$ & \NA\\
{Trigger}            &    & \\
-- Muon channel & $1.3\%$  & $4\%$\\
-- Electron  channel & $1\%$ & \NA \\
{Energy and momentum (muon channel)}          &  & \\
-- Signal mean  &   $0.04\%$ & $0.08\%$\\
-- Signal resolution &  $4\%$ & $5\%$\\
{Energy (electron channel)}         &  & \\
-- Signal mean &  $0.15\%$ & \NA\\
-- Signal resolution &  $4\%$ & \NA\\
{Jet energy scale}  & & \\
-- Muon channel & $2.5\%$ & $3.8\%$\\
-- Electron channel & $2.7\%$ & \NA \\
{Jet energy resolution}  & & \\
-- Muon channel & $0.3\%$ & $0.7\%$\\
-- Electron channel & $0.3\%$ & \NA\\
\end{tabular}
}
\end{table*}

Based on the fit bias studies, the uncertainty in the background
estimation due to the chosen functional form is assumed to be
negligible.  Furthermore, to combine the
$\PH\to\cPZ\gamma\to\ell\ell\gamma$ and
$\PH\to\gamma^*\gamma\to\mu\mu\gamma$ channels, uncertainties from
theoretical sources, integrated luminosity, object identification,
$\RNINE$ reweighting, jet energy correction and resolution are
considered to be correlated across the categories.

\begin{figure}[hbtp!]
  \centering
   \includegraphics[width=0.8\textwidth]{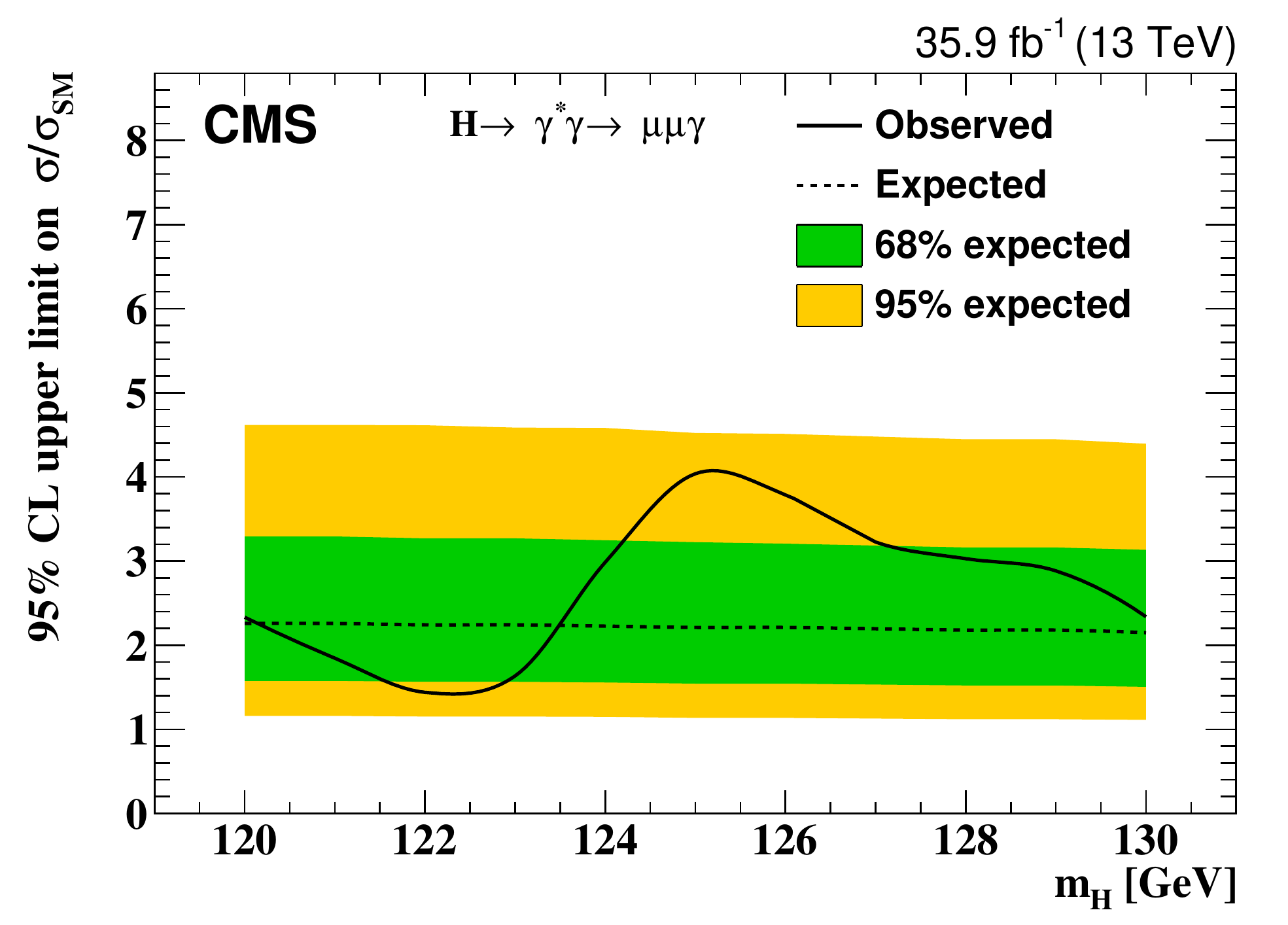} \\
    \includegraphics[width=0.8\textwidth]{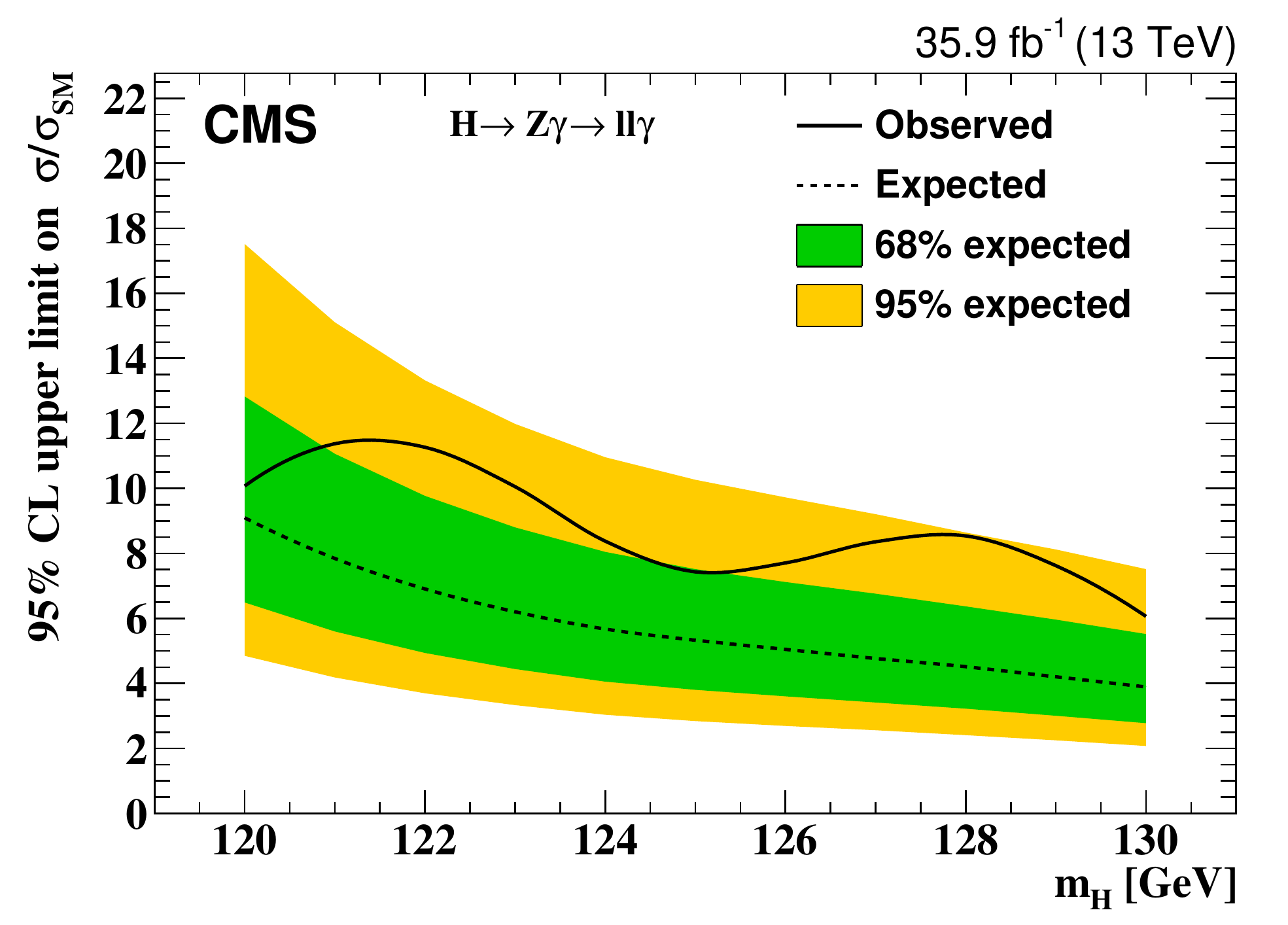}
    \caption{
Exclusion limit, at 95\% \CL, on the cross section of the $\PH\to\gamma^*\gamma\to\mu\mu\gamma$
 process (upper plot) and the $\PH\to\cPZ\gamma\to\ell\ell\gamma$ process (lower plot)
relative to the SM prediction, as a function of the Higgs boson mass.
    \label{fig:lim-dalitz}}
\end{figure}

The expected and observed exclusion limits at 95\% \CL for the process
$\PH\to\gamma^*\gamma\to\mu\mu\gamma$ are shown in
Fig~\ref{fig:lim-dalitz}.  The expected limits are between 2.1 and 2.3
times the SM cross section and the observed limit varies between about
1.4 and 4.0 times the SM cross section. The limits are calculated at
1\GeV intervals in the mass range of $120 < m_\PH < 130\GeV$.
Figure~\ref{fig:lim-dalitz} also shows the combined limit for the
$\PH\to\cPZ\gamma\to\ell\ell\gamma$ channel. The expected exclusion
limits at 95\% \CL are between 3.9 and 9.1 times the SM cross section
and the observed limit varies between about 6.1 and 11.4 times the SM
cross section.

\begin{figure}[hbtp]
  \centering
   \includegraphics[width=0.8\textwidth]{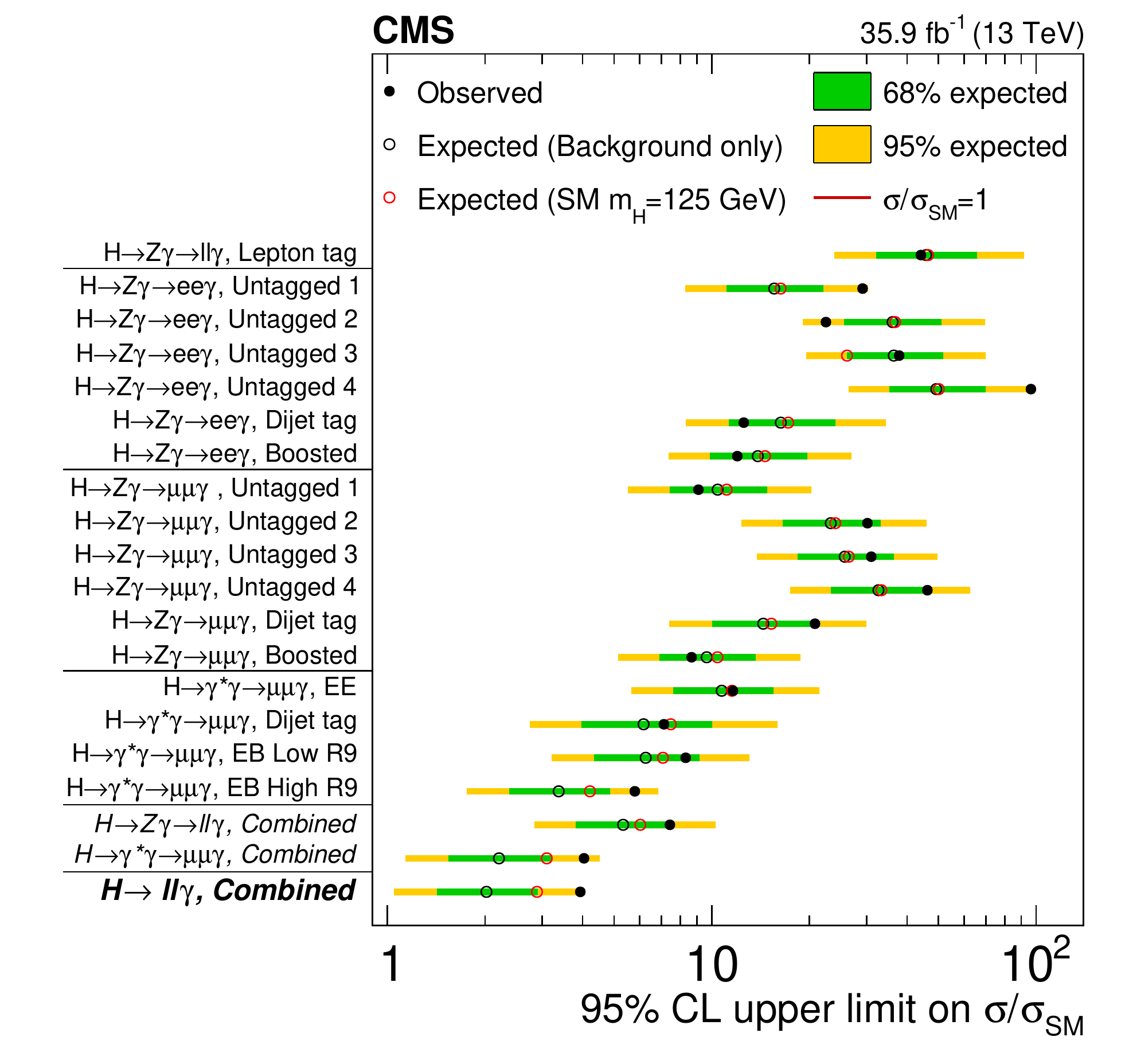}
    \caption{
Exclusion limit, at 95\% \CL, on the cross section of
$\PH\to\ell\ell\gamma$ relative to the SM prediction,
for an SM Higgs boson of $m_\PH=125\GeV$. The upper limits of each analysis category,
as well as their combinations, are shown. Black full (empty) circles show the
observed (background only expected) limit. Red circles show the expected upper limit
assuming an SM Higgs boson decaying to $\ell\ell\gamma$ decay channel.
    \label{fig:lim-combo125}}
\end{figure}

Finally, Fig.~\ref{fig:lim-combo125} shows the expected limit for each
category and the combined limit for both channels for
$m_\PH=125\GeV$. The combined observed (background only expected)
limit is 3.9 (2.0) for a 125\GeV Higgs boson decaying to
$\ell\ell\gamma$.  The same figure shows the combined expected limit
of 2.9, assuming an SM Higgs boson with $m_{\PH} = 125\GeV$, decaying
to the $\ell\ell\gamma$ channel.  After combining both analyses,
$\PH\to\gamma^*\gamma\to\mu\mu\gamma$ and
$\PH\to\cPZ\gamma\to\ell\ell\gamma$ and considering the
background-only hypothesis, the observed $p$-value at
$m_\PH=125\GeV$ is $0.02$, which corresponds to about two standard
deviations.  The combined expected $p$-value for an SM Higgs
boson at $m_\PH=125\GeV$ is $0.16$, corresponding to a significance of
around one standard deviation.

\section{Summary}
\label{sec:summary}

A search is performed for a standard model (SM) Higgs boson decaying
into a lepton pair and a photon.  This final state has contributions
from Higgs boson decays to a $\cPZ$ boson and a photon
($\PH\to\cPZ\gamma\to\ell\ell\gamma$,$\ell=\Pe$ or $\mu$),
or to two photons, one of which has an internal conversion
into a muon pair
($\PH\to\gamma^{*}\gamma\to\mu\mu\gamma$).
The analysis is performed using a data set from $\Pp\Pp$ collisions at
a center-of-mass energy of 13\TeV, corresponding to an integrated
luminosity of 35.9\fbinv. No significant excess above the
expected background is found.  Limits on the Higgs boson
production cross section times the corresponding branching fractions
are set.  The expected exclusion limits at 95\% confidence level
are about 2.1--2.3 (3.9--9.1) times the SM cross section in the
$\PH\to\gamma^*\gamma\to\mu\mu\gamma$
($\PH\to\cPZ\gamma\to\ell\ell\gamma$) channel in the mass range from
120 to 130\GeV, and the observed limit varies between about 1.4 and 4.0
(6.1 and 11.4) times the SM cross section. Finally, the
$\PH\to\gamma^*\gamma\to\mu\mu\gamma$ and
$\PH\to\cPZ\gamma\to\ell\ell\gamma$ analyses are combined for
$m_\PH=125\GeV$, obtaining an observed (expected) 95$\%$ confidence level upper limit
of 3.9 (2.0) times the SM cross section.

\clearpage

\begin{acknowledgments}

We congratulate our colleagues in the CERN accelerator departments for the excellent performance of the LHC and thank the technical and administrative staffs at CERN and at other CMS institutes for their contributions to the success of the CMS effort. In addition, we gratefully acknowledge the computing centres and personnel of the Worldwide LHC Computing Grid for delivering so effectively the computing infrastructure essential to our analyses. Finally, we acknowledge the enduring support for the construction and operation of the LHC and the CMS detector provided by the following funding agencies: BMWFW and FWF (Austria); FNRS and FWO (Belgium); CNPq, CAPES, FAPERJ, FAPERGS, and FAPESP (Brazil); MES (Bulgaria); CERN; CAS, MoST, and NSFC (China); COLCIENCIAS (Colombia); MSES and CSF (Croatia); RPF (Cyprus); SENESCYT (Ecuador); MoER, ERC IUT, and ERDF (Estonia); Academy of Finland, MEC, and HIP (Finland); CEA and CNRS/IN2P3 (France); BMBF, DFG, and HGF (Germany); GSRT (Greece); NKFIA (Hungary); DAE and DST (India); IPM (Iran); SFI (Ireland); INFN (Italy); MSIP and NRF (Republic of Korea); LAS (Lithuania); MOE and UM (Malaysia); BUAP, CINVESTAV, CONACYT, LNS, SEP, and UASLP-FAI (Mexico); MBIE (New Zealand); PAEC (Pakistan); MSHE and NSC (Poland); FCT (Portugal); JINR (Dubna); MON, RosAtom, RAS, RFBR, and NRC KI (Russia); MESTD (Serbia); SEIDI, CPAN, PCTI, and FEDER (Spain); Swiss Funding Agencies (Switzerland); MST (Taipei); ThEPCenter, IPST, STAR, and NSTDA (Thailand); TUBITAK and TAEK (Turkey); NASU and SFFR (Ukraine); STFC (United Kingdom); DOE and NSF (USA).

\hyphenation{Rachada-pisek} Individuals have received support from the Marie-Curie programme and the European Research Council and Horizon 2020 Grant, contract No. 675440 (European Union); the Leventis Foundation; the A. P. Sloan Foundation; the Alexander von Humboldt Foundation; the Belgian Federal Science Policy Office; the Fonds pour la Formation \`a la Recherche dans l'Industrie et dans l'Agriculture (FRIA-Belgium); the Agentschap voor Innovatie door Wetenschap en Technologie (IWT-Belgium); the F.R.S.-FNRS and FWO (Belgium) under the ``Excellence of Science - EOS" - be.h project n. 30820817; the Ministry of Education, Youth and Sports (MEYS) of the Czech Republic; the Lend\"ulet (``Momentum") Programme and the J\'anos Bolyai Research Scholarship of the Hungarian Academy of Sciences, the New National Excellence Program \'UNKP, the NKFIA research grants 123842, 123959, 124845, 124850 and 125105 (Hungary); the Council of Science and Industrial Research, India; the HOMING PLUS programme of the Foundation for Polish Science, cofinanced from European Union, Regional Development Fund, the Mobility Plus programme of the Ministry of Science and Higher Education, the National Science Center (Poland), contracts Harmonia 2014/14/M/ST2/00428, Opus 2014/13/B/ST2/02543, 2014/15/B/ST2/03998, and 2015/19/B/ST2/02861, Sonata-bis 2012/07/E/ST2/01406; the National Priorities Research Program by Qatar National Research Fund; the Programa Estatal de Fomento de la Investigaci{\'o}n Cient{\'i}fica y T{\'e}cnica de Excelencia Mar\'{\i}a de Maeztu, grant MDM-2015-0509 and the Programa Severo Ochoa del Principado de Asturias; the Thalis and Aristeia programmes cofinanced by EU-ESF and the Greek NSRF; the Rachadapisek Sompot Fund for Postdoctoral Fellowship, Chulalongkorn University and the Chulalongkorn Academic into Its 2nd Century Project Advancement Project (Thailand); the Welch Foundation, contract C-1845; and the Weston Havens Foundation (USA).

\end{acknowledgments}

\bibliography{auto_generated}
\cleardoublepage \appendix\section{The CMS Collaboration \label{app:collab}}\begin{sloppypar}\hyphenpenalty=5000\widowpenalty=500\clubpenalty=5000\vskip\cmsinstskip
\textbf{Yerevan Physics Institute, Yerevan, Armenia}\\*[0pt]
A.M.~Sirunyan, A.~Tumasyan
\vskip\cmsinstskip
\textbf{Institut f\"{u}r Hochenergiephysik, Wien, Austria}\\*[0pt]
W.~Adam, F.~Ambrogi, E.~Asilar, T.~Bergauer, J.~Brandstetter, E.~Brondolin, M.~Dragicevic, J.~Er\"{o}, A.~Escalante~Del~Valle, M.~Flechl, R.~Fr\"{u}hwirth\cmsAuthorMark{1}, V.M.~Ghete, J.~Hrubec, M.~Jeitler\cmsAuthorMark{1}, N.~Krammer, I.~Kr\"{a}tschmer, D.~Liko, T.~Madlener, I.~Mikulec, N.~Rad, H.~Rohringer, J.~Schieck\cmsAuthorMark{1}, R.~Sch\"{o}fbeck, M.~Spanring, D.~Spitzbart, A.~Taurok, W.~Waltenberger, J.~Wittmann, C.-E.~Wulz\cmsAuthorMark{1}, M.~Zarucki
% ...
\vskip\cmsinstskip
% ...
1:  Also at Vienna University of Technology, Vienna, Austria\\
% ...
\end{sloppypar}
\end{document}